\documentclass[twocolumn, nofootinbib, longbibliography, amsmath, amssymb, citeautoscript, prl, aps, superscriptaddress]{revtex4-2}

\usepackage[english]{babel}
\usepackage[utf8]{inputenc}
\usepackage{lmodern, soul}
\usepackage{dcolumn}
\usepackage{amssymb, amsmath}
\usepackage{graphicx}
\graphicspath{{figures/}} 
\usepackage{bm}
\usepackage{xcolor}
\usepackage{nicefrac}
\usepackage[%
colorlinks=true,
urlcolor=blue,
linkcolor=blue,
citecolor=blue
]{hyperref}
\usepackage{cleveref}
\usepackage{verbatim}
\usepackage{mathtools}
\usepackage{epstopdf}
\usepackage{bbold}
\usepackage{comment}
\usepackage{placeins}
\usepackage{enumitem}
\usepackage{booktabs}	 
\usepackage{multirow}
\usepackage{microtype}
\usepackage{braket}
\usepackage{physics}
\usepackage{tabularx}
\usepackage{cellspace} 
\usepackage{dsfont} 
\usepackage{lineno}
\usepackage[normalem]{ulem}

\begin{document}
\title{Dominant orbital magnetization in the prototypical altermagnet MnTe}

\author{Chao Chen Ye}
\email{c.chen.ye@rug.nl}
\affiliation{Zernike Institute for Advanced Materials, University of Groningen, Nijenborgh 3, 9747 AG Groningen, The Netherlands}
\author{Karma Tenzin}
\affiliation{Zernike Institute for Advanced Materials, University of Groningen, Nijenborgh 3, 9747 AG Groningen, The Netherlands}
\author{Jagoda S{\l}awi{\'n}ska}
\email{jagoda.slawinska@rug.nl}
\affiliation{Zernike Institute for Advanced Materials, University of Groningen, Nijenborgh 3, 9747 AG Groningen, The Netherlands}
\author{Carmine Autieri}
\email{autieri@magtop.ifpan.edu.pl}
\affiliation{International Research Centre MagTop, Institute of Physics, Polish Academy of Sciences, Aleja Lotnik\'ow 32/46, PL-02668 Warsaw, Poland}
\affiliation{SPIN-CNR, UOS Salerno, IT-84084 Fisciano (SA), Italy}

\begin{abstract}
Altermagnetism is an unconventional form of antiferromagnetism characterized by momentum-dependent spin polarization of electronic states and zero net magnetization, arising from specific crystalline symmetries. In the presence of spin-orbit coupling (SOC), altermagnets can exhibit finite net magnetization and anomalous Hall effect (AHE), phenomena typically associated with ferromagnets. As AHE is closely linked to weak ferromagnetism, clarifying its microscopic origin in terms of spin and orbital contributions to magnetization is crucial for interpreting experiments and designing altermagnetic devices. In this work, we use density functional theory to explore the intrinsic spin and orbital magnetization of the magnetic ground state of the prototypical altermagnet $\alpha$-MnTe. We find that SOC induces weak ferromagnetism through spin canting, accompanied by a slight in-plane rotation of the N\'eel vector. Notably, we identify a significant net orbital magnetization of 0.176~$\mu_B$ per unit cell oriented along the $z$-axis, whereas the spin magnetization in the same direction is much smaller, at only 0.002~$\mu_B$. By varying the chemical potential, we show that the spin magnetization can be tuned through hole doping, whereas the orbital magnetization remains robust against carrier concentration changes. These results highlight the important role of orbital magnetization and establish its relevance for orbital-based phenomena in altermagnets. 
\end{abstract}

\maketitle

\section{Introduction}
\label{sec:introduction}

Altermagnetism is a recently identified form of antiferromagnetism, characterized by nonrelativistic spin splitting of electronic bands in momentum space despite a vanishing net magnetization, and has been proposed in several material classes~\cite{doi:10.7566/JPSJ.88.123702,Yuan2023,smejkal_2022, smejkal_2022_2, guo_2023, tamang2024newlydiscoveredmagneticphase, cuono_2023, PhysRevMaterials.8.L051401, PhysRevB.108.115138}. It has attracted attention for its potential in highly efficient spin-current generation~\cite{gonzalez_2021}, optical altermagnetic switch~\cite{devita2025opticalswitchinglayeredaltermagnet}, anomalous Hall effect (AHE) ~\cite{Takagi2025,sheoran2025spontaneousanomaloushalleffect}, giant tunneling magnetoresistance~\cite{smejkal_2022_3,sun2025altermagnetizingfeseliketwodimensionalmaterials}, antiferromagnetic spintronics~\cite{ding_2021, ding_2023}, spin caloritronics~\cite{zhou_2024} and Josephson junctions~\cite{ouassou_2023}, among other applications. Its existence relies on the condition that the electronic charge of spin-up (down) atoms maps onto spin-down (up) 
ones through symmetry operations involving rotations - whether proper or improper, symmorphic or nonsymmorphic - but not by pure inversion or translation. This symmetry constraint gives rise to characteristic $d$-wave, $g$-wave, or $i$-wave patterns of nonrelativistic spin polarization in momentum space~\cite{smejkal_2022}. Such patterns are absent in antiferromagnets that preserve combined time-reversal (T) and spatial symmetries of inversion (P) or lattice translation ($\tau$)~\cite{Xiao2021,Cheong2024}. Although spin-orbit coupling (SOC) is often neglected in descriptions of altermagnetism, it can induce finite magnetization that influences the system's behavior.

Materials with a small net magnetization induced by SOC are called \textit{weak ferromagnets}~\cite{dzyaloshinsky_1958}. SOC generates effective antisymmetric interactions between spins~\cite{mazurenko2005weak}, most notably the staggered Dzyaloshinskii–Moriya interaction (DMI)~\cite{autieri2312staggered,PhysRevB.111.064401}, which is closely linked to orbital effects~\cite{moriya_1960}. Several recent works aim to classify the different types of SOC-driven antisymmetric interactions~\cite{roig2024quasisymmetryconstrainedspinferromagnetism,cheong2025altermagnetismclassification,Cheong2024,mcclarty_2024,schiff2024collinearaltermagnetslandautheories,Jo2024gtensor, solovyev2025altermagnetismweakferromagnetism}. In weak ferromagnets, the magnitude of the net magnetic moment is determined by the interplay of magnetic exchange interactions, magnetocrystalline anisotropy, and SOC-induced effects. Most commonly, it appears via spin canting - a small deviation from perfect antiferromagnetic alignment. Since magnetic exchange generally dominates over SOC, the resulting canting angles are usually small, often less than 1$^\circ$~\cite{autieri2312staggered}, unless the magnetic atoms are heavy transition metals~\cite{srdjan}.


\begin{figure}[htb!]
	\centering
	\includegraphics[width=0.49\linewidth]{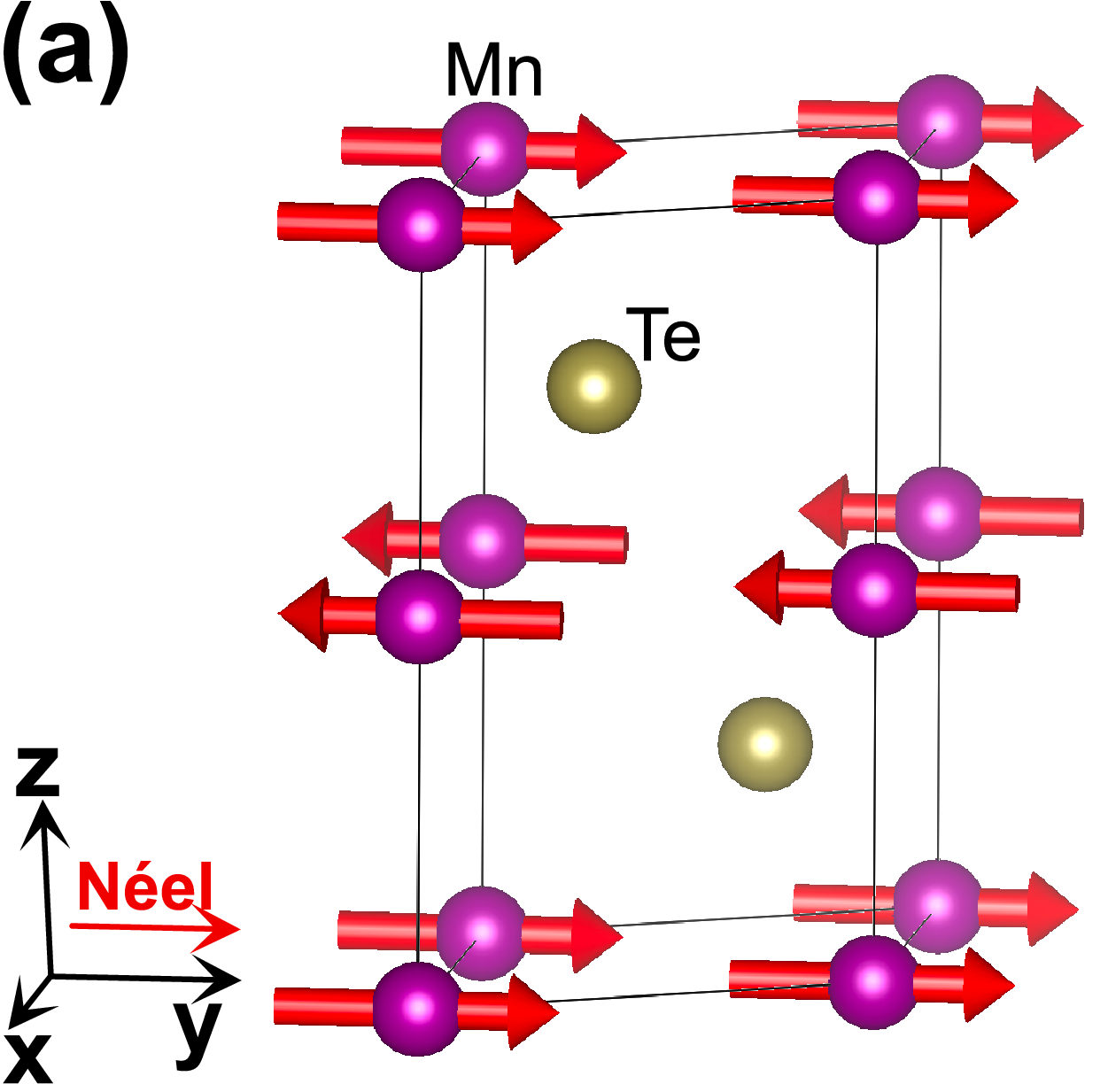}
	\includegraphics[width=0.49\linewidth]{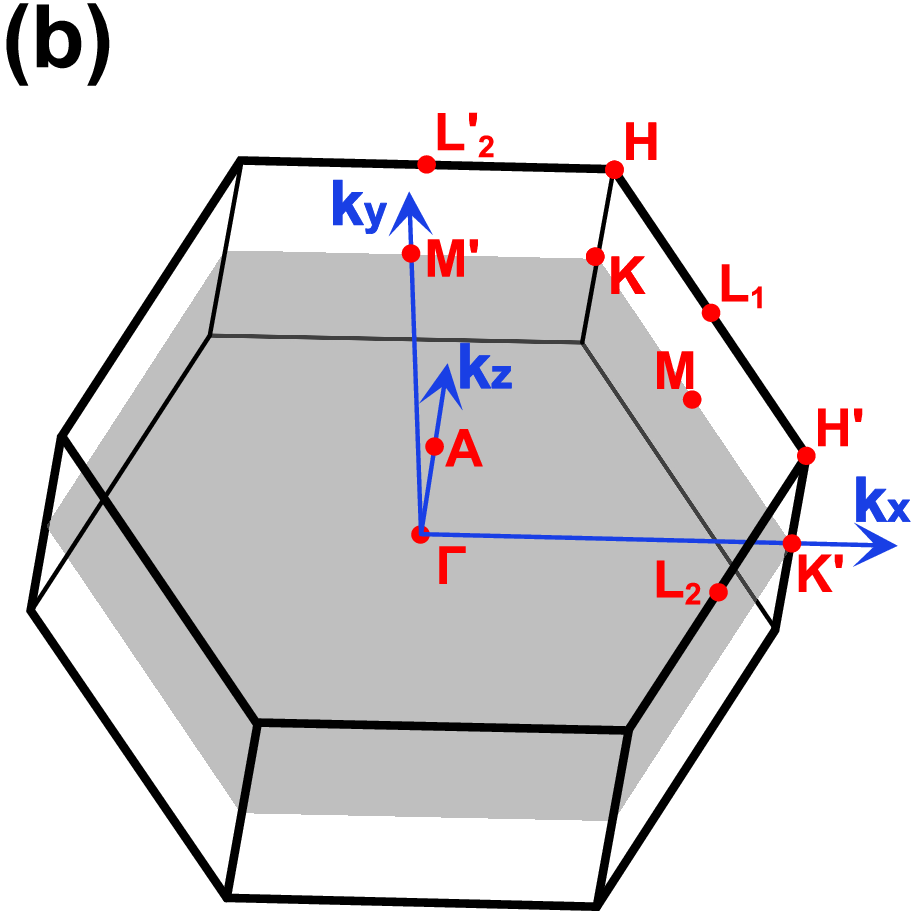}
	\includegraphics[width=0.7\linewidth]{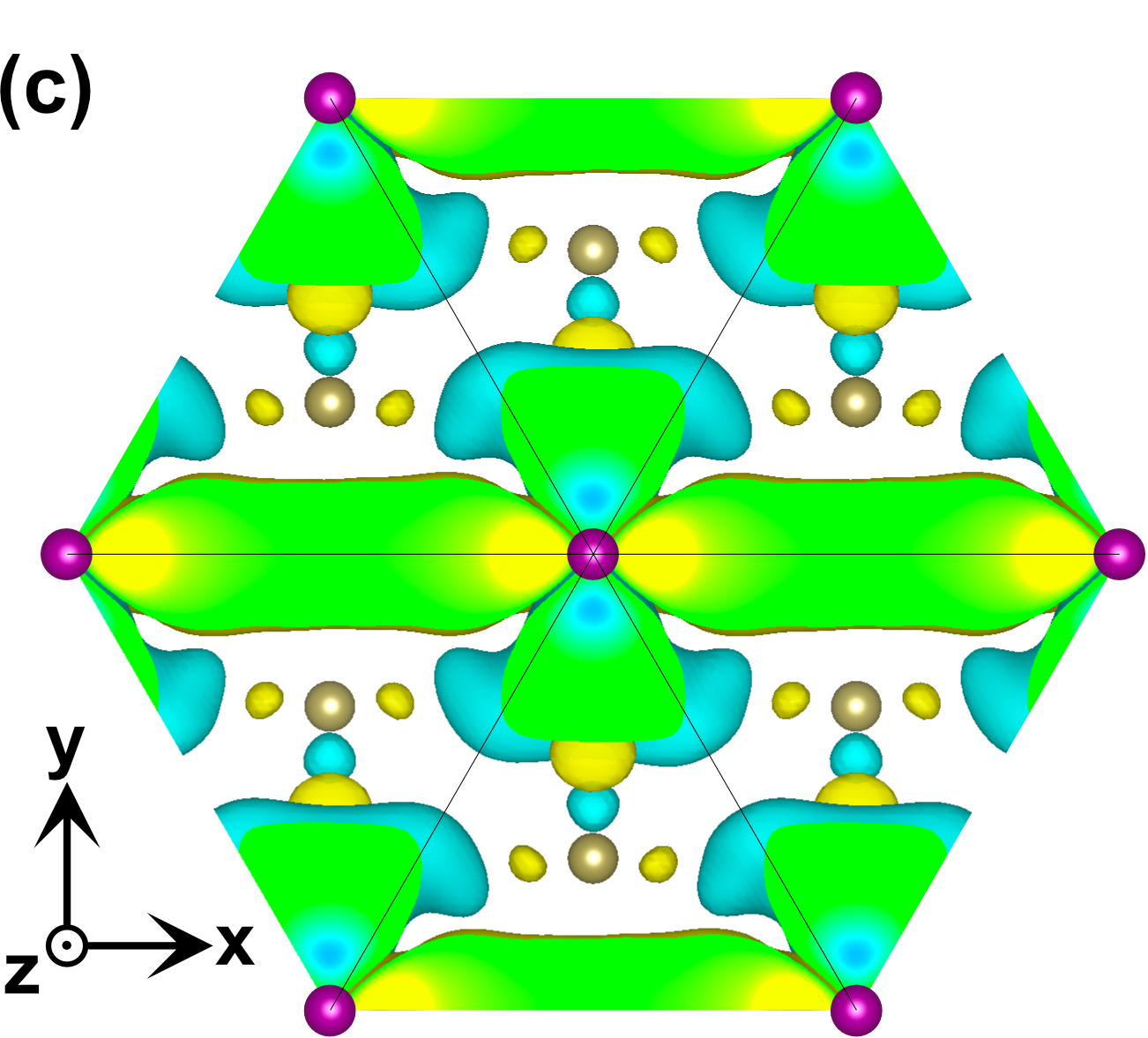} 
	\caption{\textbf{Structure and charge density of MnTe.} \textbf{(a)} Crystal structure with the magnetic moment orientation along the $y$-axis. \textbf{(b)} The corresponding hexagonal Brillouin zone. \textbf{(c)} Difference between the charge densities of the N\'eel vector along the $y$-axis and $x$-axis for the isosurface $1.5\cdot 10^{-6} \; e/${\AA}$^3 $. The blue and yellow colors indicate the positive and negative values of the charge differences, respectively. 
    The breaking of $\qty{6^{\pm}_{001}\vert 0,0,\frac{1}{2}}$ and $\qty{3^{\pm}_{001}\vert 0}$ symmetries is observed, while the inversion $\qty{\bar{1}\vert0}$  and $\qty{2_{001}\vert 0,0,\frac{1}{2}}$ symmetries are preserved.}
	\label{fig:atomic_stucture_bz}
\end{figure} 

$\alpha$-MnTe is a prototypical altermagnet, known for exhibiting one of the largest nonrelativistic spin splittings of electronic bands~\cite{krempasky2024altermagnetic}. It has been extensively studied both experimentally and theoretically~\cite{kriegner2016multiple, yin_2019, mori2020reversible, mazin2024origin,PhysRevMaterials.8.104407,GonzalezBetancourt2024, lee_2024, liu_2024, belashchenko_2025, Takahashi2025}. The symmetries responsible for spin splitting also generate the SOC-driven antisymmetric spin interactions~\cite{autieri2312staggered}, though their magnitudes are not directly correlated, as the former depends on kinetic parameters and the latter on the SOC strength and order~\cite{roig2024quasisymmetryconstrainedspinferromagnetism}. In MnTe, SOC leads to net spin magnetization, which makes it a weak ferromagnet~\cite{kluczyk_2024}.  One of the key factors behind its emergence is the magnetocrystalline anisotropy, which must be small to not suppress the spin canting. In MnTe, this anisotropy was recently estimated to be $K_2 = 40~\mu$eV~\cite{dzian2025antiferromagneticresonancealphamnte}, where $K_2 S_z^2$ is the energy cost of spins acquiring a $z$-component. In addition, AHE was predicted and observed in MnTe~\cite{kluczyk_2024, PhysRevLett.130.036702}, 
 and the large observed signals may indicate that both spin and orbital contributions play a significant role ~\cite{PhysRevB.90.045202,PhysRevB.105.235142,PhysRevLett.133.236602, liu2025different}.
Therefore, a detailed understanding of orbital magnetization is essential for unveiling microscopic origin of weak ferromagnetism and for guiding future studies of orbital-driven phenomena in altermagnets ~\cite{doi:10.1126/sciadv.adn3662,sobral2024fractionalizedaltermagnetsneighboringaltermagnetic}.

In this work, we perform density functional theory (DFT) simulations to quantitatively investigate spin and orbital magnetization in the altermagnetic MnTe. Our goal is to estimate these two contributions with respect to doping and to unveil the importance of orbital magnetization. We start by exploring the system's symmetries through the analysis of charge density differences and relativistic spin-resolved electronic structures. Moreover, based on the calculations of spin and orbital magnetization for different doping levels, we reveal that both are present in the insulating phase. We show that orbital magnetization is aligned primarily along the $z$-direction, with the net orbital contribution two orders of magnitude stronger than the spin counterpart. In contrast to the net spin magnetization, which changes with hole doping, the orbital magnetization remains nearly constant over a large energy range. 

\section{Structural and electronic properties}

\subsection{Symmetries of the system}

In this subsection, we introduce the notation for the crystal structure and we summarize the electronic properties. $\alpha$-MnTe is an intrinsic $p$-type doped magnetic semiconductor with a N\'eel temperature of around 310~K~\cite{uchida1956magnetic, banewicz1961high, ozawa1966effect, ferrer2000temperature} and a large band gap~\cite{ferrer2000temperature, osumi2024}. The term $\alpha$ refers to the hexagonal nickeline crystal structure, built by alternating layers of Mn and Te atoms, described by the $P6_{3}/mmc$ (\#194) space group, which only considers the symmetries relating atomic positions in the crystallographic lattice. Introducing magnetic moments onto the Mn atoms and assuming the N\'eel vector along the $y$-axis, imposes additional symmetry restrictions which lower the symmetry to the magnetic space group $Cm^{\prime}c^{\prime}m$ (\#63.462)~\cite{kriegner2017}. This configuration is depicted in Fig.~\ref{fig:atomic_stucture_bz}(a). In terms of the Seitz notation, where the prime symbol indicates the preservation of the time-reversal operation, we identified eight symmetries in Cartesian coordinates that are preserved even in the presence of SOC-induced magnetic canting. These are:
$\qty{1\vert 0}$, 
$\qty{\bar{1}\vert0}$, 
$\qty{2_{001}\vert 0,0,\frac{1}{2}}$, 
$\qty{m_{001}\vert 0,0,\frac{1}{2}}$,
$\qty{2^{\prime}_{010}\vert 0,0,\frac{1}{2}}$, 
$\qty{m^{\prime}_{010}\vert 0,0,\frac{1}{2}}$, 
$\qty{2^{\prime}_{100}\vert 0}$ and 
$\qty{m^{\prime}_{100}\vert 0}$, 
as determined using \texttt{Spglib}~\cite{Togo31122024, spglib2}.

Note that the symmetry reduction from $P6_{3}/mmc$ to $Cm^{\prime}c^{\prime}m$ transforms the crystal from a primitive hexagonal to a base-centered orthorhombic cell, which in turn alters the corresponding Brillouin zone (BZ)~\cite{Cracknell1969}. Despite this change, both BZs share the shape of a hexagonal prism; the primary difference lies in the high-symmetry points. In this work, we adopt the notation associated with the primitive hexagonal lattice, shown in Fig.~\ref{fig:atomic_stucture_bz}(b), which includes all relevant high-symmetry $k$-points. When introducing the N\'eel vector along the $y$-axis, the crystal symmetry is lowered and only the twofold rotational symmetry remains. As a consequence, we have other inequivalent $\rm{L}$ points beyond $\rm{L_1}$ and $\rm{L_2}$, which we denote as $\rm{L_2'}$.
The symmetry breaking becomes even more apparent in real space. By computing the charge density difference between configurations with the N\'eel vector aligned along the $x$- and $y$-axes, we observe the loss of $\qty{6^{\pm}_{001}\vert 0,0,\frac{1}{2}}$ symmetry, with only the nonsymmorphic $\qty{2_{001}\vert 0,0,\frac{1}{2}}$ operation preserved. This confirms the breaking of the threefold rotational symmetry $\qty{3^{\pm}_{001}\vert 0}$\cite{yin_2019, mazin2024origin}. The resulting charge density difference is visualized in Fig.\ref{fig:atomic_stucture_bz}(c).

\begin{figure*}[htb!]
\centering
\includegraphics[width=0.31\linewidth]{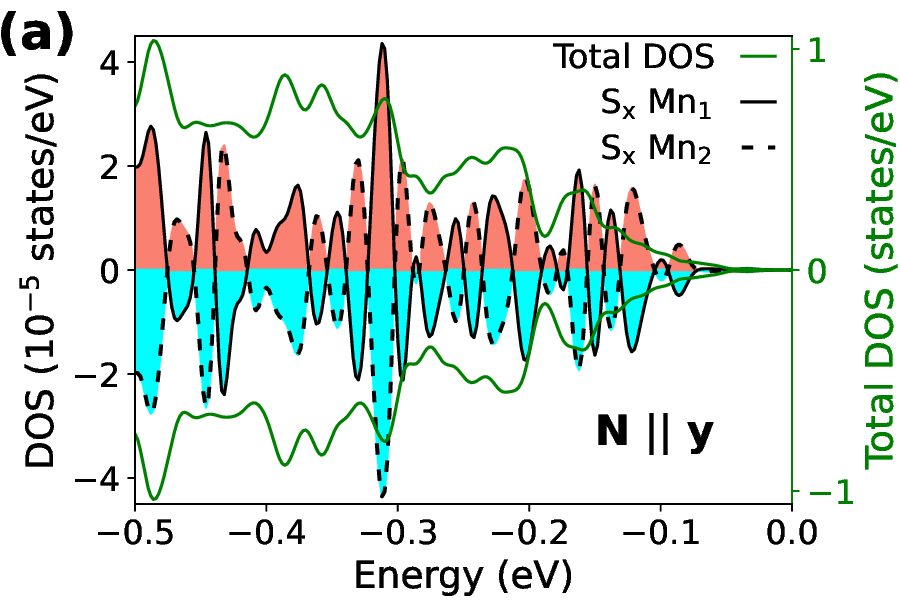} \quad
\includegraphics[width=0.31\linewidth]{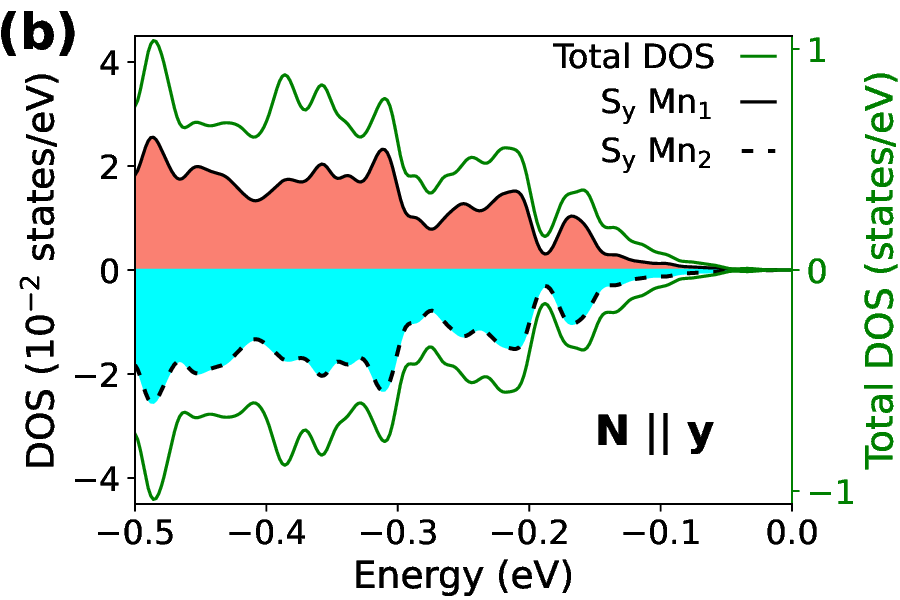} \quad
\includegraphics[width=0.31\linewidth]{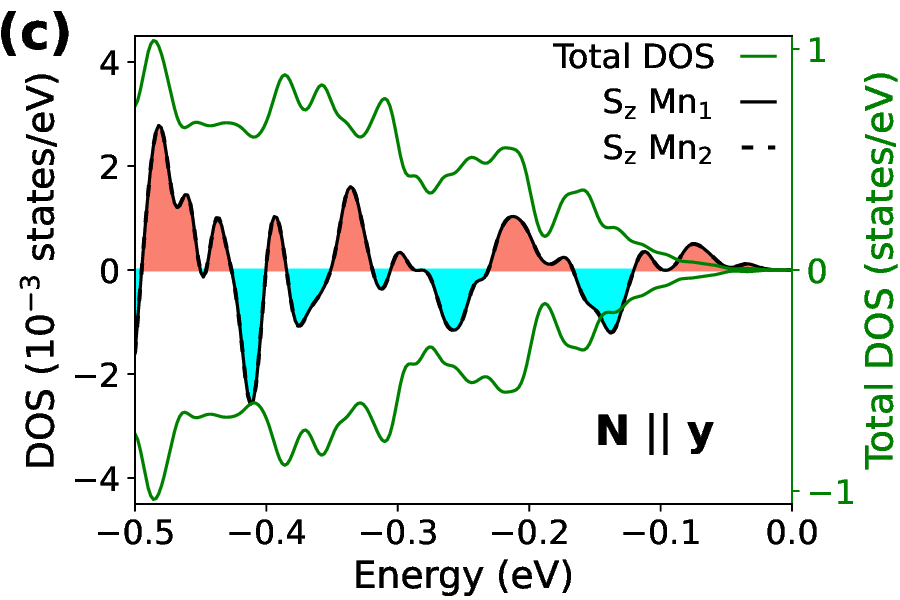}
\caption{\textbf{Relativistic spin density of states in the valence bands for MnTe with N\'eel vector along the $y$-axis (\textbf{N}~$||$~\textbf{y}).} \textbf{(a)} $S_x$ is the component responsible for the slight rotation of the N\'eel vector in the $xy$-plane. \textbf{(b)} $S_y$ is the main component of the N\'eel vector. \textbf{(c)} $S_z$ component is responsible for the weak ferromagnetism along the $z$-axis. Positive and negative spin contributions are shown in red and blue, respectively. The total DOS is overlaid in green in all panels.
}
\label{fig:spindensity}
\end{figure*}

\subsection{Spin density of states}
\label{sec:spindensity}

We start with calculating the site-projected, energy-resolved magnetization density, denoted as $\vb{S}^{\text{Mn}_1}$ and $\vb{S}^{\text{Mn}_2}$ for the projections onto Mn$_1$ and Mn$_2$, respectively. We refer to these quantities hereafter as the spin density of states. The  N\'eel vector is defined as $\vb{N} = \frac{1}{2}(\vb{S}^{\text{Mn}_1} - \vb{S}^{\text{Mn}_2}$), while the spin magnetization per unit cell is given by $\vb{M}^{\text{spin}} = \vb{S}^{\text{Mn}_1} + \vb{S}^{\text{Mn}_2}$. The calculated spin density of states is reported in Fig.~\ref {fig:spindensity}. Note that only the spin angular momentum is considered; the orbital contribution to the magnetization will be addressed in the next section.

The dominating spin component, $S_y$, shown in Fig.~\ref{fig:spindensity}(b), is orders of magnitude larger than the others and gives the main contribution to the total density of states (DOS). The spin components $S_y$ of Mn$_{1}$ and Mn$_{2}$ are equal and opposite. In addition, we plot the total DOS, including all atoms, in green; we can observe that the relativistic spin density of states of the $S_y$ component resembles the total DOS. The spin components $S_x$ of the Mn$_{1}$ and Mn$_{2}$ are also equal and opposite, as shown in Fig.~\ref{fig:spindensity}(a). Based on the calculated relativistic spin density of states, we can see that there is a slight in-plane rotation of the N\'eel vector. When building the free energy with N\'eel vector $N = (N_x, N_y, N_z)$ and magnetization $M = (M_x, M_y, M_z)$ as order parameter, the energy describing the spin-orbit coupling effect $M_z N_y (3N_x^2 -N_y^2)$~\cite{roig2024quasisymmetryconstrainedspinferromagnetism} does not contain first-order terms in the N\'eel vector as the DMI interaction. It implies that the spin canting in MnTe arises from a higher-order SOC-driven interaction. This rotation of the N\'eel vector depends on the chemical potential and, therefore, can be manipulated by doping. 

The $S_z$ component of spin density of states is equal and has the same sign for both Mn$_1$ and Mn$_2$, as can be seen by the overlap of solid and dashed lines in Fig.~\ref{fig:spindensity}(c); this component generates the weak ferromagnetism. We can also observe that the spin density of $S_z$ oscillates as a function of the energy. The oscillation is related to the fact that the weak ferromagnetism is proportional to the SOC-driven antisymmetric interaction. This behavior is captured by the SOC-driven parameter $D_2$ (see Supplementary Material), which varies with the energy as the DMI~\cite{PhysRevB.102.224427, koretsune_2018}. Consistently, a toy model for altermagnets has shown that such oscillations of DMI with band filling are an intrinsic property in these systems~\cite{beutier_2017}.


To summarize, we have the following relations for the integrated spin components $S_x^{Mn_{1}}=-S_x^{Mn_{2}}$, $S_y^{Mn_{1}}=-S_y^{Mn_{2}}$, $S_z^{Mn_{1}}=S_z^{Mn_{2}}$ with the order relation being $S_y^{Mn_{1}} >>  S_z^{Mn_{1}} >> S_x^{Mn_{1}}$. Although the value of $S_x^{Mn_{1}}$ is extremely small to give sizable effects, it is interesting to note that it is not zero by symmetry. Since the integrals of $S_x$ and $S_y$ are zero by symmetry for every value of the Fermi level, in the rest of the paper, we will focus on the $S_z$ component projected on the band structure.

\begin{figure}[tb!]
	\centering
	\includegraphics[width=\linewidth]{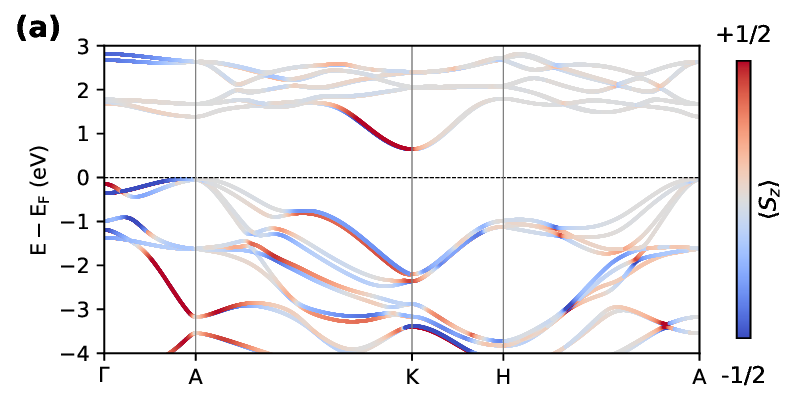}
	\includegraphics[width=\linewidth]{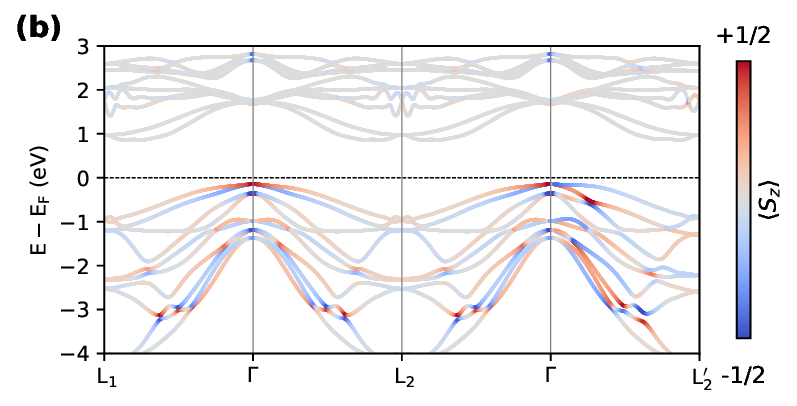}
	\caption{\textbf{Relativistic band structure with the $S_z$ component of the spin texture indicated by the color scale on the side.} We plot the band structure along the $k$-path \textbf{(a)} $\rm{\Gamma-A-K-H-A}$ and \textbf{(b)} $\rm{L_1-\Gamma-L_2-\Gamma-L'_2}$. The Fermi energy is set at the VBM.}
	\label{fig:spin_resolved_bands}
\end{figure} 

\subsection{Altermagnetic electronic structure with SOC}
\label{sec:electronic}
Relativistic altermagnetic band structures calculated along different high-symmetry lines are presented in Fig.~\ref{fig:spin_resolved_bands}. These calculations reveal signatures of both magnetic ordering, associated with the local magnetization around Mn atoms, and strong SOC originating from Te atoms \cite{ferrer2000temperature, faria2023sensitivity, mazin2024origin, Takahashi2025}. Although Fig.~\ref{fig:spin_resolved_bands}(a) indicates an indirect band gap of 0.68 eV between the $\rm{A}$ and $\rm{K}$ points, smaller than the experimental value of approximately 1.27 eV \cite{ferrer2000temperature}, we note that this discrepancy will not affect our conclusions, as MnTe is intrinsically $p$-type doped and our analysis is focused on the valence band region. Additionally, we anticipate that the valence band maximum (VBM) is not at the high-symmetry point $\rm{A}$, instead, it is located along the $\rm{\Gamma-K'}$ line (see the next section and Fig. \ref{fig:morb_path}).

In terms of the orbital character of the electronic states, it is well established that the orbitals near the valence band maximum and conduction band minimum (CBM) in MnTe are mainly composed of ``Te-$5p$ + Mn-$3d$" and ``Mn-$3d$ + Mn-$4s$" states, respectively~\cite{allen_1977, ferrer2000temperature, faria2023sensitivity, PhysRevB.103.115209}. The bands at the high-symmetry points $\Gamma$ and A are fourfold degenerate. More specifically, the top of the valence band at the A point is doubly degenerate, dominated by Te $p_x$–$p_y$ orbital character, while the singly degenerate $p_z$-like state lies farther below the Fermi level. We note that although the nonrelativistic band structure hosts the largest spin splitting along the $\rm{\Gamma-L}$ directions, it does not strictly hold anymore in the relativistic case. 
 
The coloring of the bands in Fig~\ref{fig:spin_resolved_bands} represents the $S_z$ projection of the spin texture and reveals that this component is consistently present across the entire band structure when the N\'eel vector is oriented along the $y$-axis~\cite{krempasky2024altermagnetic}. Focusing on the $\rm{L_1-\Gamma-L_2}$ $k$-path, we find that the projection of the main $S_y$ component is opposite along the two paths, as detailed in the Supplementary Material. In contrast, the $S_z$ component - responsible for weak ferromagnetism - remains the same along both $\rm{L_1-\Gamma}$ and $\rm{\Gamma-L_2}$ in Fig.~\ref{fig:spin_resolved_bands}(b). Notably, our band structure is asymmetric between $\rm{L_2}$  and $\rm{L_2'}$, which further confirms the sixfold symmetry breaking~\cite{krempasky2024altermagnetic}. 

\begin{figure*}[htb!]
	\centering
	\includegraphics[width=\linewidth]{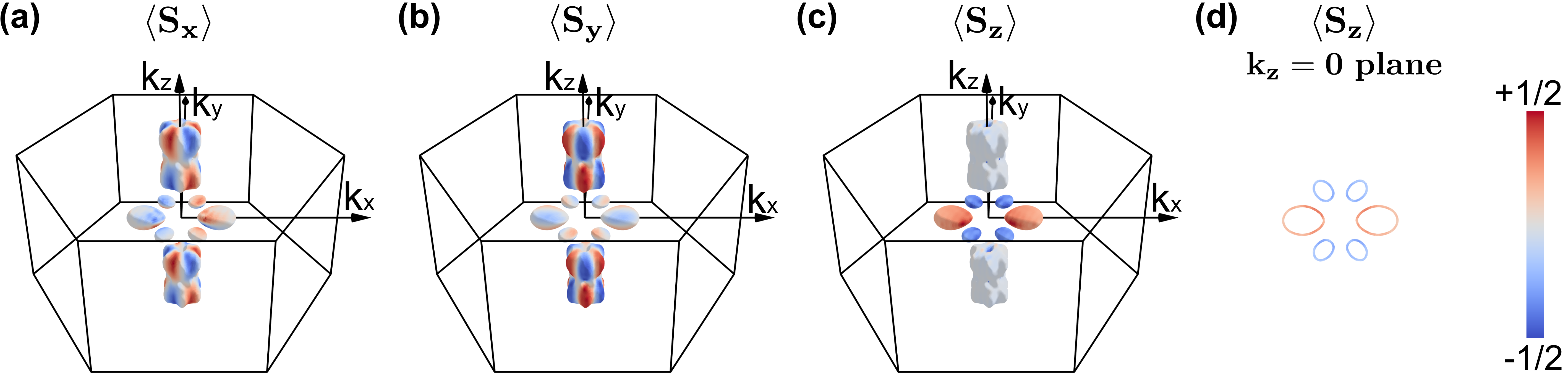}
	\caption{\textbf{Spin-projected isosurfaces with the Fermi level in the valence band}. Isosurface $E-E_{F} \approx -0.10$ eV colored by the component of the spin texture \textbf{(a)} $S_x$, \textbf{(b)} $S_y$ and \textbf{(c)} $S_z$. \textbf{(d)} The isoenergy contours for the $k_z = 0$ plane together with the overlaid spin texture component $S_z$.}
	\label{fig:spin_texture}
\end{figure*}

Finally, we examine the Fermi pockets across the Brillouin zone under a small hole doping, as shown in Fig.~\ref{fig:spin_texture}. These pockets correspond to the isosurface at $E - E_F \approx -0.10$ eV, located below the VBM. The spin texture components $S_x$ and $S_y$ are shown in Fig.~\ref{fig:spin_texture}(a) and (b), respectively. The most pronounced feature, at first sight, is the $g$-wave spin polarization pattern, visible in the larger pockets near the BZ edges, where the spin direction changes every 60$^\circ$, reminiscent of the nonrelativistic case. This pattern is absent in Fig.~\ref{fig:spin_texture} (c), which shows the $S_z$ component. While the alternating $g$-wave structure is clearly observed for the in-plane components $S_x$ and $S_y$, no higher rotational symmetry beyond a twofold rotation can be inferred from the dominant red lobes in the $S_z$ texture. A slice view of Fig.~\ref{fig:spin_texture}(c) showing the plane $k_z = 0$ is depicted in Fig.~\ref{fig:spin_texture}(d). We note that the Fermi pockets with $+S_z$ are much larger compared to those with $- S_z$, but this size difference would be hard to observe experimentally~\cite{krempasky2024altermagnetic}. Additionally, we highlight the $S_z$ contributions along the $k_z$-direction, displayed inside the white color Fermi pockets in Fig.~\ref{fig:spin_texture}(c).\\ 



\subsection{Spin and orbital magnetization}
\label{sec:spin_and_orbital_magnetization}

We now numerically estimate the total magnetization per unit cell, $\vb{M}=\vb{M}^{spin}+\vb{M}^{orb}$, consisting of both spin and orbital contributions. We only consider intrinsic mechanisms responsible for the magnetization; the extrinsic mechanisms, such as structural and magnetic defects, will not be considered in the present paper~\cite{Bugajewski:2025_prb}. In particular, the orbital magnetization is expected to be relevant for materials with large SOC and it is present in systems with large DMI~\cite{moriya_1960, adamantopoulos_2024, mazin2024origin, autieri2312staggered}; however, few results have been reported for altermagnets. The calculations of spin and orbital magnetization are evaluated as the sum of contributions from all Bloch states over the entire Brillouin zone~\cite{xiao_2005, thonhauser_2005, ceresoli_2006,  xiao_2010, vanderbilt_2012, johansson2024theory, burgos_2024}, using the following equations:
\begin{widetext}
\begin{equation}
	\begin{split}
		&\vb{M}^{spin} = \sum_{n} \int_{BZ}  \frac{\dd \vb{k} }{\qty(2 \pi)^3} \, f_n( \vb{k}) \vb{M}^{spin}_n (\vb{k}) = \frac{g_s \mu_B }{ \qty(2 \pi)^3}  \sum_{n} \int_{BZ} \dd \vb{k}   f_n( \vb{k}) \expval{\hat{\vb{S}}}{\psi_{n \vb{k}}} \; , \\
		&\vb{M}^{orb} = \sum_{n} \int_{BZ} \frac{\dd \vb{k} }{\qty(2 \pi)^3} \, f_n( \vb{k}) \vb{M}^{orb}_n (\vb{k}) =  \frac{e}{2 \hbar \qty(2 \pi)^3} \Im \sum_{n} \int_{BZ} \dd \vb{k}   f_n( \vb{k}) \expval{ \cross \qty(\hat{H}_{\vb{k}} + E_{n \vb{k} }  - 2 E_F ) }{\pdv{u_{n \vb{k}} }{ \vb{k} } }    \; , \\
	\end{split}
	\label{eq:magnetization}
\end{equation}
\end{widetext}
where $g_s$ is the spin $g$-factor, $\mu_B$ is the Bohr magneton, $e$ is the elementary charge, and $ f_n (\vb{k})$ is an electronic distribution function. The operator $\hat{\vb{S}}$ represents the spin, $\hat{H}_{\vb{k}}$ is the Bloch Hamiltonian, and  $\ket{\psi_{n \vb{k}} }= e^{i \vb{k} \cdot \vb{r}} \ket{u_{n \vb{k}} } $ are Bloch eigenstates with eigenvalues $ E_{n \vb{k} }$. The spin expectation values have already been discussed in the previous sections; now, we focus on orbital magnetization calculated along different paths in the BZ. 

\begin{figure}[htb!]
	\centering
	\includegraphics[width=\linewidth]{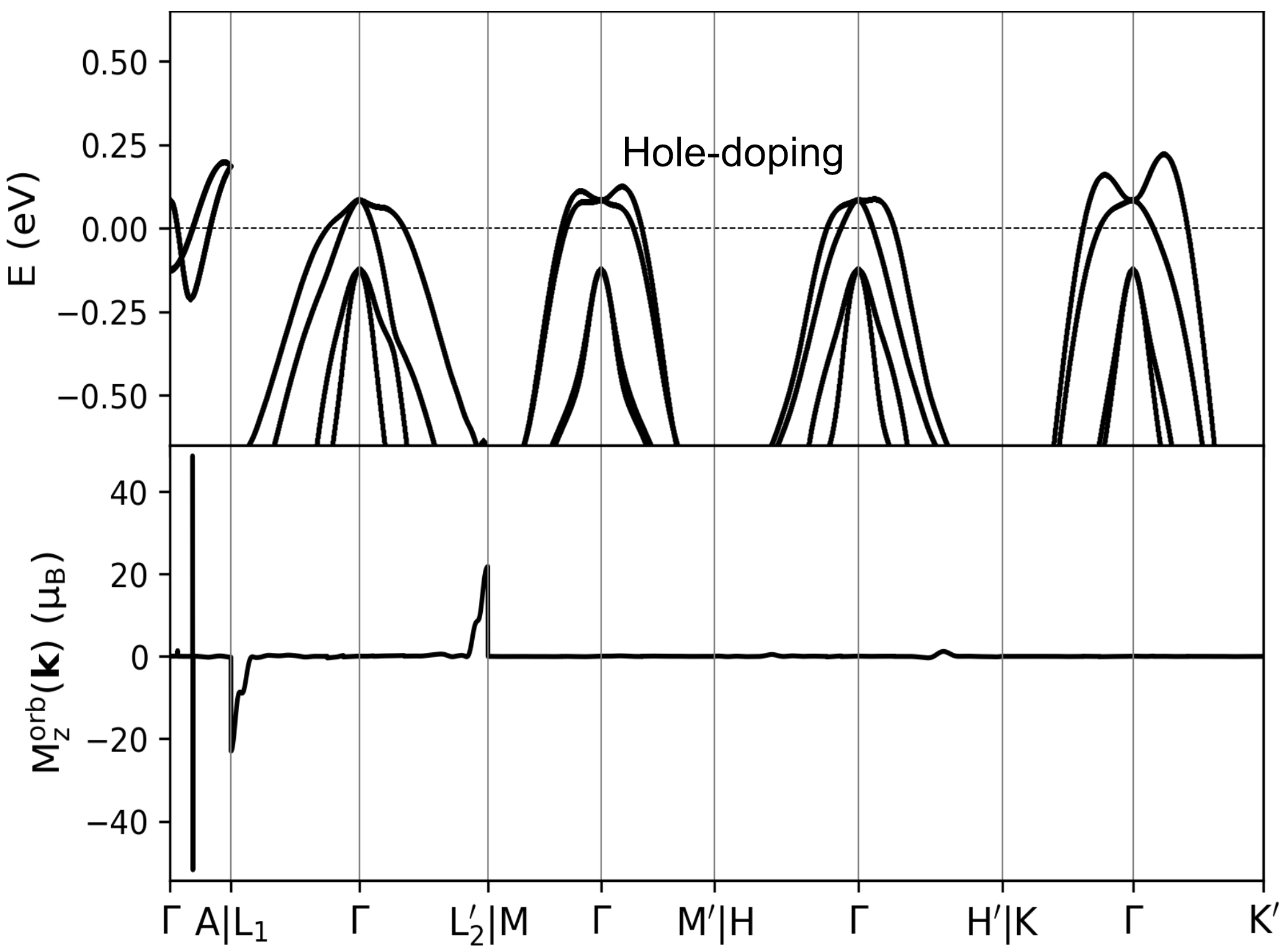}
	%
	%
	\caption{\textbf{Calculated k-resolved orbital magnetization $M^{orb}_z (\vb{k})$ for the Fermi level in the valence band.} Orbital magnetization $M^{orb}_z (\vb{k})$ along $k$-paths for $E_F = E_{VBM} -0.23 $ eV, that is for the Fermi level shifted to the valence band. Note that $M^{orb}_x (\vb{k})$ and $M^{orb}_y (\vb{k})$ are negligible compared to $M^{orb}_z (\vb{k})$, as shown in the Supplementary Material. 
    }
	\label{fig:morb_path}
\end{figure}

Figure~\ref{fig:morb_path} shows the $k$-resolved orbital magnetization evaluated with the chemical potential shifted to simulate sizable hole doping. For reference, the upper panel displays the corresponding band structure near the selected chemical potential to facilitate the interpretation of the orbital magnetization behavior along various $k$-paths. We observe that the most prominent orbital magnetization peak in the valence band appears along the $\rm{\Gamma - A}$ path, i.e., along the $k_z$-axis. The $M^{\text{orb}}_x(\vb{k})$ and $M^{\text{orb}}_y(\vb{k})$ net components are negligible, therefore, $M^{\text{orb}}_z(\vb{k})$ is the only net component (see Supplementary Material). The value of $M^{\text{orb}}_z(\vb{k})$ along $\rm{\Gamma - A}$ becomes zero once the valence band is filled upon low-level electron-doping (see Supplementary Material), therefore, the k-points along $\rm{\Gamma - A}$ do not give a net contribution to the orbital magnetization. 

To emphasize the importance of orbital magnetization in MnTe, we note that the bcc Fe exhibits peaks that are at least one order of magnitude smaller~\cite{vanderbilt_2012} than the peaks displayed in Fig.~\ref{fig:morb_path}. Additionally, we identify a negative peak at $\rm{L_1}$ and a positive peak $\rm{L_2'}$ which originate from bands located deeply in the occupied part of the spectrum - this is a consequence of the cumulative nature of the orbital magnetization. We recall that there are four $\rm{L}$ points and two $\rm{L'}$ points; the summed contribution yields a negative orbital magnetization. 
Further contributions to the orbital magnetization will arise from k-points that are not located along the high-symmetry lines.


\begin{figure}[htb!]
	\centering
	\includegraphics[width=\linewidth]{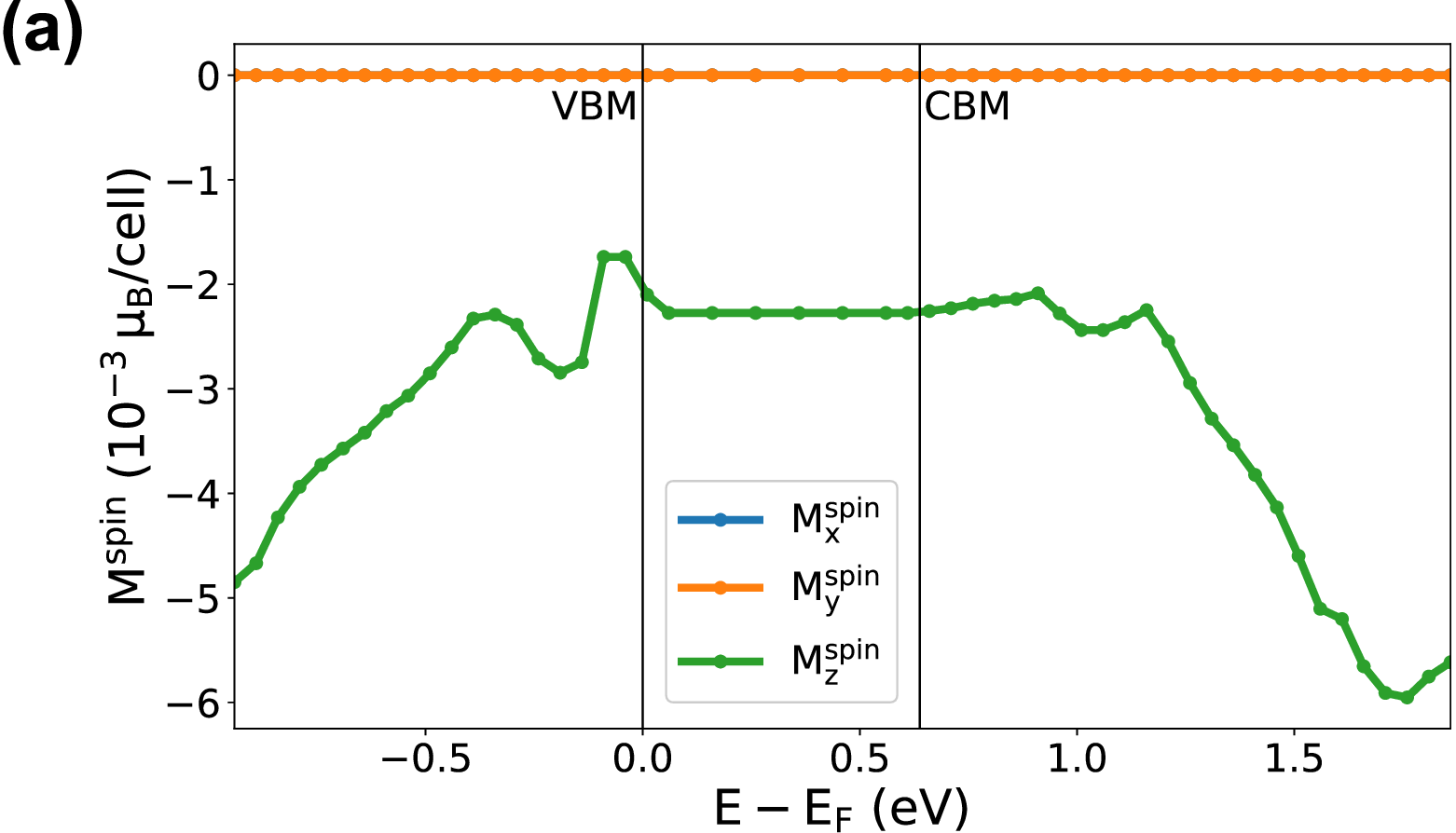}
	\includegraphics[width=\linewidth]{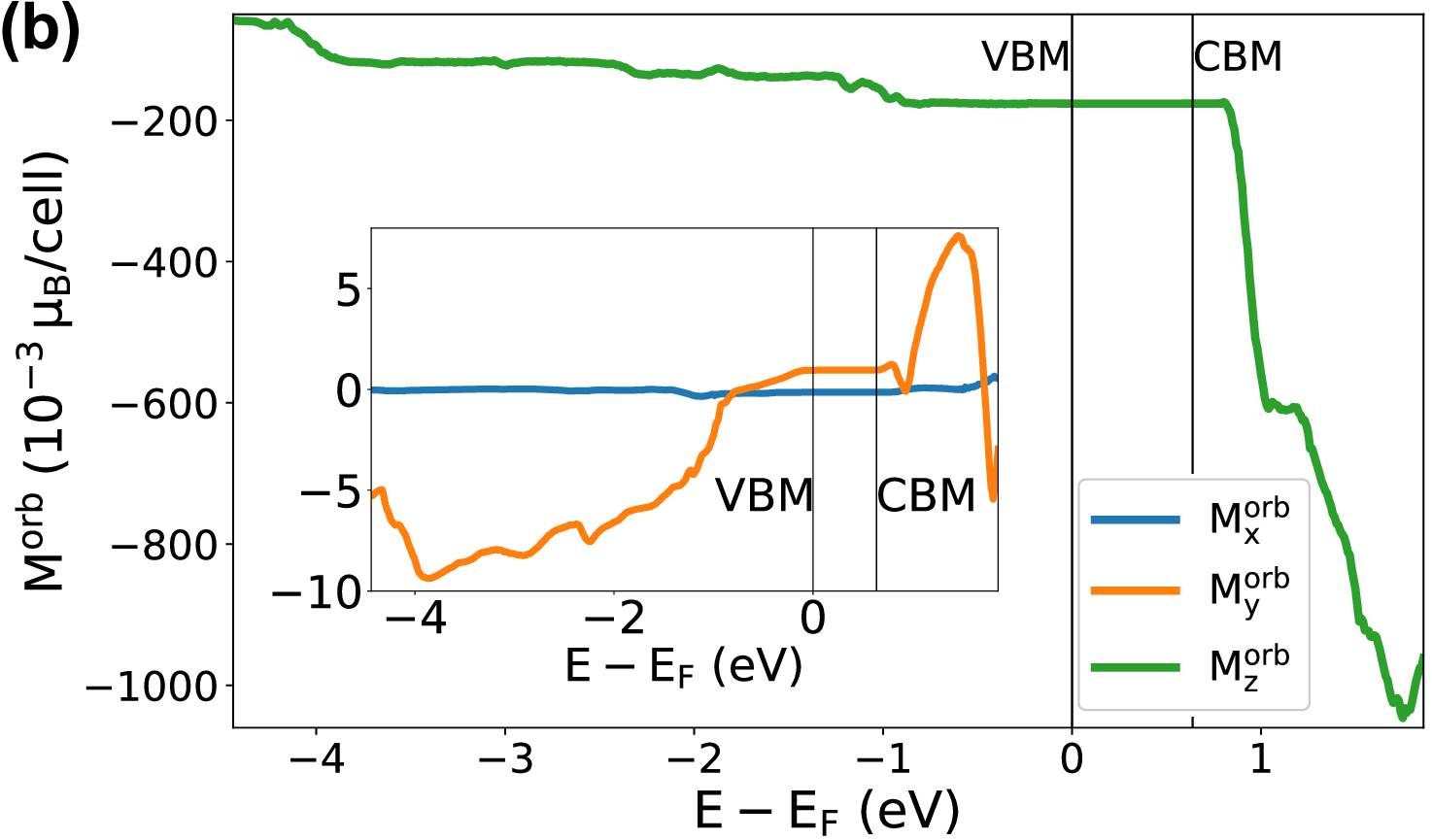}
	\caption{\textbf{Net spin and orbital magnetizations of MnTe in the low-energy window.} \textbf{(a)} Spin magnetization components M$_x^{spin}$ and M$_y^{spin}$ are exactly zero as expected from symmetry. \textbf{(b)} Orbital magnetization component. M$_z^{orb}$ is the dominant contribution, while M$_x^{orb}$ and M$_y^{orb}$ are shown in the inset. The $y$-axis of the inset figure displays a different range compared to the main figure. The two vertical lines indicate the VBM and CBM.}
	\label{fig:spin_and_orbital_magnetizations}
\end{figure}

Finally, we evaluate the net spin and orbital magnetization as a function of the chemical potential, governed by Eq.~\eqref{eq:magnetization}, which relates total magnetization to the Fermi level. The results for the spin magnetization are shown in Fig.~\ref{fig:spin_and_orbital_magnetizations}(a). Due to the symmetry of the SOC-driven antisymmetric interaction, the net spin magnetization has vanishing components in the $x$- and $y$-directions, i.e., $M^{\text{spin}}_x = M^{\text{spin}}_y = 0$. In contrast, the $z$-component remains finite across all values of the chemical potential and does not vanish even within the band gap. This implies that altermagnets can exhibit weak ferromagnetism even in the absence of doping. Since $M^{\text{spin}}_z$ depends on the filling, we can infer that the hole doping influences the magnetic canting. 

However, the weak ferromagnetism is strongly suppressed by the presence of a large band gap. Since it arises from the spin-nonconserving nature of spin-orbit coupling, which enables mixing between spin-up and spin-down states on the same atomic site, a large energy gap between the spin-up and spin-down manifolds compared to the SOC strength makes this mixing negligible (see Supplementary Material). This results in an almost vanishing imbalance between spin-up and spin-down components on each site. As a consequence, weak ferromagnetism is expected to be suppressed in wide-gap, high-spin, undoped systems. This suppression is confirmed numerically: at the Fermi level, where the valence band is fully occupied, the net spin magnetization reaches a near-minimum in MnTe. Our DFT calculations yield a total spin magnetization of approximately 0.002~$\mu_B$ per unit cell. Each Mn atom contributes about 0.001~$\mu_B$ along the $z$-axis, compared to a total spin moment of 4~$\mu_B$, from which we obtain that the canting angle (described in Supplementary Material) is $\overline{\theta}$=0.01$^\circ$.

%

The components of the net orbital magnetization calculated as a function of the chemical potential are reported in Fig.~\ref{fig:spin_and_orbital_magnetizations}(b). Similarly to spin magnetization, the orbital magnetization is constant inside the band gap due to the absence of electronic states. In this case, however, $M^{orb}_x  \simeq  0$ is not exactly zero but on the order of $10^{-5} \, \mu_B$. Similarly, $M^{orb}_y  \simeq  0$ but its magnitude is around $10^{-4} \, \mu_B$. The dominant term is $M^{\text{orb}}_z$, reaching approximately $-0.176 , \mu_B$ per unit cell within the band gap. Notably, the orbital and spin magnetization have the same direction, but the orbital contribution is two orders of magnitude larger. This indicates the crucial role of orbital magnetization in MnTe, whose magnitude exceeds that of bcc Fe, despite MnTe having a total magnetization an order of magnitude smaller~\cite{meyer_1961, vanderbilt_2012, hanke_2016, kubler_1981}. Notably, this finding is in line with the previous theoretical studies, which also reported orbital magnetization dominating over spin contributions in weak ferromagnets~\cite{Jo2024gtensor}. 

Experimentally, the net magnetization - including both spin and orbital parts - is in the range of $10^{-4}$ to $10^{-3}~\mu_B$ per Mn atom~\cite{kluczyk_2024, PhysRevLett.132.176701}, significantly lower than our theoretical prediction ${M_z}^{\text{tot}} = {M_z}^{\text{spin}} + {M_z}^{\text{orb}} \approx -0.178~\mu_B$. This discrepancy can be attributed to the presence of altermagnetic domains, which tend to compensate each other, leading to a reduced observed net magnetization~\cite{Amin2024, takegami2025circulardichroismresonantinelastic}. 

In summary, along the $z$-axis, the orbital contribution clearly dominates, confirming its major role for shaping weak ferromagnetism. These results indicate the need to incorporate orbital contributions, via Eq.~\eqref{eq:magnetization}, for a complete and accurate description of altermagnetic materials. More detailed studies are still needed to reveal the exact origin of the large orbital magnetization; possible hypotheses include the contributions from Weyl points discovered in altermagnets with the NiAs structure \cite{Li2025_Weyl}.

\section{Conclusions}
\label{sec:conclusions}
In conclusion, we have performed \textit{ab initio} density functional theory calculations for altermagnet $\alpha$-MnTe to investigate its spin and orbital magnetization. 

We find that both spin and orbital magnetization vectors are aligned along the $z$-axis, perpendicular to the N\'eel vector direction. While spin canting leads to weak ferromagnetism, the resulting spin magnetization is small, approximately 0.002~$\mu_B$ per unit cell. The net spin magnetization is suppressed in MnTe by multiple factors: the higher-order SOC term, the wide band gap, and the large magnetic exchange. In contrast, the intrinsic orbital magnetization is substantial, reaching 0.176~$\mu_B$ per unit cell, and represents the dominant contribution to the net magnetization. Our calculated magnetization per unit cell is larger than the experimental value; this calls for a careful examination of altermagnetic domains and extrinsic magnetization contributions~\cite{Bugajewski:2025_prb}.

Finally, our results show that a finite magnetization persists in the insulating phase of $\alpha$-MnTe, although it is suppressed near the band gap, while hole doping enables tuning of the spin magnetization. In contrast, the orbital magnetization remains nearly constant over a wide energy range - up to 0.75~eV below the valence band maximum - demonstrating its robustness against carrier doping. 

These results show the importance of orbital magnetization in altermagnets and suggest that concepts explored in spin-based altermagnetic systems may be extended to orbital phenomena. This opens new opportunities for both spin- and orbitronics as well as designing orbitronic functionalities in altermagnetic materials. 

\section{Computational details}
\label{sec:method}

We performed DFT calculations using \texttt{Vienna Ab initio Simulation Package} (VASP)~\cite{vasp_1993, vasp_1996, kresse_1996_2}. The kinetic energy cutoff for the plane wave basis was set to 350~eV and the generalized gradient approximation of Perdew, Burke, and Ernzerhof was used for the exchange-correlation functional~\cite{perdew_1996}. The Hubbard $U$ for the Mn-$3d$ orbitals was added~\cite{liechtenstein_1995}, with the Coulomb repulsion $U = 4$ eV and the Hund coupling $J_H = 0.97 $~eV~\cite{antropov_2014, kriegner2017}. The energy convergence threshold was set to $2\cdot 10^{-5}$~eV. 

The structure was modeled by a hexagonal unit cell with lattice parameters equal to $a=b=4.134 $ {\AA} and $c=6.652 $ {\AA}~\cite{kriegner2017}. The Brillouin zone was sampled using a mesh of $20 \cross 20 \cross 12$ $k$-points centered at $\Gamma$. For the spin density of states, we used a $28 \cross 28 \cross 20$ $k$-points grid. The spin-orbit coupling was taken into account in all the calculations and the symmetries were switched on. 

As a postprocessing step, we used Wannier90~\cite{pizzi2020} to calculate spin-polarized band structures, orbital magnetization along high-symmetry lines in the Brillouin zone, as well as the net spin and orbital magnetization as a function of chemical potential. The k-grid used for the Wannier90 calculations was 100 $\times$ 100 $\times$ 64. The orbital magnetization remained unchanged when increasing the Wannier90 k-grid, whereas variations of about 10\% were observed when changing the DFT k-grid. The radius within which the magnetization is calculated is 2.500 atomic units, corresponding to 1.323 {\AA}. 

In addition, to corroborate these results, we performed parallel calculations using Quantum Espresso \cite{qe1, qe2} with comparable parameters at both the DFT and Wannier levels, obtaining consistently similar values of the orbital magnetization as a function of chemical potential.

The post-processing code \texttt{PAOFLOW}~\cite{paoflow_2018, paoflow_2021} was used to evaluate spin-polarized Fermi pockets at different values of hole doping shown. 

\section{Data Availability}
\label{sec:data_availability}
The data that support the findings of this study will be openly available at DataVerseNL XXX [repository link to be added at the proof stage].

\section{Acknowledgments}
\label{sec:acknowledgments}

The authors thank T. Dietl,  M. Mostovoy, M. Sawicki, K. V{\'y}born{\'y}, J. \v{Z}elezn{\'y}, V. V. Volobuev, A. Kazakov and B.J.M. van Dijk for useful discussions.
We acknowledge the research program “Materials for the Quantum Age” (QuMat) for financial support. This program (registration number 024.005.006) is part of the Gravitation program financed by the Dutch Ministry of Education, Culture and Science (OCW).
C. A. was supported by the Foundation for Polish Science project “MagTop” no.~FENG.02.01-IP.05-0028/23, co-financed by the European Union from the funds of Priority 2 of the European Funds for a Smart Economy Program 2021–2027 (FENG). C.A. acknowledges support from PNRR MUR project PE0000023-NQSTI. J.S. acknowledges the Rosalind Franklin Fellowship from the University of Groningen. The calculations were carried out on the Dutch national e-infrastructure with the support of SURF Cooperative (EINF-10658).

\section{Author contributions}
\label{sec:contributions}
C.C.Y. performed all calculations and wrote the initial draft of the manuscript. K.T. performed \texttt{PAOFLOW} calculations. All the authors analyzed the data and participated in the discussions. J.S. and C.A. supervised the project.

\FloatBarrier

\bibliography{lib} 

@article{schiff2024collinearaltermagnetslandautheories,
  title = {Collinear altermagnets and their Landau theories},
  author = {Schiff, Hana and McClarty, Paul and Rau, Jeffrey G. and Romh\'anyi, Judit},
  journal = {Phys. Rev. Res.},
  volume = {7},
  issue = {3},
  pages = {033301},
  numpages = {31},
  year = {2025},
  month = {Sep},
  publisher = {American Physical Society},
}

@article{mazin2024origin,
	title = {Origin of the gossamer ferromagnetism in {MnTe}},
	author = {Mazin, I. I. and Belashchenko, K. D.},
	journal = {Phys. Rev. B},
	volume = {110},
	issue = {21},
	pages = {214436},
	numpages = {7},
	year = {2024},
	month = {Dec},
	publisher = {American Physical Society},
}

@article{ferrer2000temperature,
	title={Temperature and pressure dependence of the optical absorption in hexagonal {MnTe}},
	author={Ferrer-Roca, Ch and Segura, A and Reig, C and Munoz, V},
	journal={Phys. Rev. B},
	volume={61},
	number={20},
	pages={13679},
	year={2000},
	publisher={APS}
}

@article{faria2023sensitivity,
	title={Sensitivity of the {MnTe} valence band to the orientation of magnetic moments},
	author={Faria Junior, Paulo E and de Mare, Koen A and Zollner, Klaus and Ahn, Kyo-hoon and Erlingsson, Sigurdur I and van Schilfgaarde, Mark and V{\`y}born{\`y}, Karel},
	journal={Phys. Rev. B},
	volume={107},
	number={10},
	pages={L100417},
	year={2023},
	publisher={APS}
}

@article{kriegner2017,
	title = {Magnetic anisotropy in antiferromagnetic hexagonal {MnTe}},
	author = {Kriegner, D. and Reichlova, H. and Grenzer, J. and Schmidt, W. and Ressouche, E. and Godinho, J. and Wagner, T. and Martin, S. Y. and Shick, A. B. and Volobuev, V. V. and others},
	journal = {Phys. Rev. B},
	volume = {96},
	issue = {21},
	pages = {214418},
	numpages = {8},
	year = {2017},
	month = {Dec},
}

@article{uchida1956magnetic,
	title={Magnetic and Electrical Properties of Manganese Telluride},
	author={Uchida, Enji and Kondoh, Hisamoto and Fukuoka, Nobuo},
	journal={J. Phys. Soc. Jpn.},
	volume={11},
	number={1},
	pages={27--32},
	year={1956},
	publisher={The Physical Society of Japan}
}

@article{ozawa1966effect,
	title={Effect of pressure on the magnetic transition point of manganese telluride},
	author={Ozawa, K and Anzai, S and Hamaguchi, Y},
	journal={Phys. Lett.},
	volume={20},
	number={2},
	pages={132--133},
	year={1966},
	publisher={Elsevier}
}

@article{banewicz1961high,
	title={High temperature magnetic susceptibilities Of {MnO}, {MnSe} and {MnTe}},
	author={Banewicz, John J and Heidelberg, Robert F and Luxem, Allan H},
	journal={J. Phys. Chem.},
	volume={65},
	number={4},
	pages={615--617},
	year={1961},
	publisher={ACS Publications}
}

@article{osumi2024,
	title = {Observation of a giant band splitting in altermagnetic {MnTe}},
	author = {Osumi, T. and Souma, S. and Aoyama, T. and Yamauchi, K. and Honma, A. and Nakayama, K. and Takahashi, T. and Ohgushi, K. and Sato, T.},
	journal = {Phys. Rev. B},
	volume = {109},
	issue = {11},
	pages = {115102},
	numpages = {8},
	year = {2024},
	month = {Mar},
	publisher = {American Physical Society},
}

@article{adamantopoulos_2024,
	title={Spin and orbital magnetism by light in rutile altermagnets},
    author={Adamantopoulos, Theodoros and Merte, Maximilian and Freimuth, Frank and Go, Dongwook and Zhang, Lishu and Ležaić, Marjana and Feng, Wanxiang and Yao, Yugui and Sinova, Jairo and Šmejkal, Libor and Blügel, Stefan and Mokrousov, Yuriy},
	journal={npj Spintronics},
    volume={2},
	number={1},
	publisher={Springer Science and Business Media LLC},
	year={2024},
	month={Sep} 
}

@misc{devita2025opticalswitchinglayeredaltermagnet,
      title={Optical switching in a layered altermagnet}, 
      author={Alessandro De Vita and Chiara Bigi and Davide Romanin and Matthew D. Watson and Vincent Polewczyk and Marta Zonno and François Bertran and My Bang Petersen and Federico Motti and Giovanni Vinai and others},
      year={2025},
      eprint={2502.20010},
      archivePrefix={arXiv},
}

@article{PhysRevLett.133.236602,
  title = {In-Plane Anomalous {H}all Effect Associated with Orbital Magnetization: Measurements of Low-Carrier Density Films of a Magnetic {W}eyl Semimetal},
  author = {Nakamura, Ayano and Nishihaya, Shinichi and Ishizuka, Hiroaki and Kriener, Markus and Watanabe, Yuto and Uchida, Masaki},
  journal = {Phys. Rev. Lett.},
  volume = {133},
  issue = {23},
  pages = {236602},
  numpages = {6},
  year = {2024},
  month = {Dec},
  publisher = {American Physical Society},
}

@article{PhysRevB.102.224427,
  title = {Semirealistic tight-binding model for {D}zyaloshinskii-{M}oriya interaction},
  author = {Hajr, Ahmed and Hariri, Abdulkarim and Manchon, Guilhem and Ghosh, Sumit and Manchon, Aur\'elien},
  journal = {Phys. Rev. B},
  volume = {102},
  issue = {22},
  pages = {224427},
  numpages = {11},
  year = {2020},
  month = {Dec},
  publisher = {American Physical Society},
}

@misc{sun2025altermagnetizingfeseliketwodimensionalmaterials,
      title={Altermagnetizing the {FeSe}-like two-dimensional materials and approaching to giant tunneling magnetoresistance with {J}anus {C}r$_{4}${B}{N}({B}$_{2}$) {MB}ene electrode}, 
      author={Weiwei Sun and Mingzhuang Wang and Baisheng Sa and Shaui Dong and Carmine Autieri},
      year={2025},
      eprint={2502.03165},
      archivePrefix={arXiv},
}

@article{doi:10.7566/JPSJ.88.123702,
author = {Hayami ,Satoru and Yanagi ,Yuki and Kusunose ,Hiroaki},
title = {Momentum-Dependent Spin Splitting by Collinear Antiferromagnetic Ordering},
journal = {J. Phys. Soc. Jpn.},
volume = {88},
number = {12},
pages = {123702},
year = {2019},
}

@article{PhysRevB.108.115138,
  title = {Interplay between altermagnetism and nonsymmorphic symmetries generating large anomalous {H}all conductivity by semi-{D}irac points induced anticrossings},
  author = {Fakhredine, Amar and Sattigeri, Raghottam M. and Cuono, Giuseppe and Autieri, Carmine},
  journal = {Phys. Rev. B},
  volume = {108},
  issue = {11},
  pages = {115138},
  numpages = {8},
  year = {2023},
  month = {Sep},
  publisher = {American Physical Society},
}

@article{PhysRevLett.130.036702,
  title = {Spontaneous Anomalous {H}all Effect Arising from an Unconventional Compensated Magnetic Phase in a Semiconductor},
  author = {Gonzalez Betancourt, R. D. and Zub\'a\ifmmode \check{c}\else \v{c}\fi{}, J. and Gonzalez-Hernandez, R. and Geishendorf, K. and \ifmmode \check{S}\else \v{S}\fi{}ob\'a\ifmmode \check{n}\else \v{n}\fi{}, Z. and Springholz, G. and Olejn\'{\i}k, K. and \ifmmode \check{S}\else \v{S}\fi{}mejkal, L. and Sinova, J. and Jungwirth, T. and others},
  journal = {Phys. Rev. Lett.},
  volume = {130},
  issue = {3},
  pages = {036702},
  numpages = {7},
  year = {2023},
  month = {Jan},
  publisher = {American Physical Society},
}

@article{dzian2025antiferromagneticresonancealphaMnTe,
      title={Antiferromagnetic resonance in $\alpha$-{MnTe}}, 
      author={J. Dzian and P. Kubaščík and S. Tázlarů and M. Białek and M. Šindler and F. Le Mardelé and C. Kadlec and F. Kadlec and M. Gryglas-Borysiewicz and K. P. Kluczyk and others},
  journal = {Phys. Rev. B},
  volume = {112},
  issue = {2},
  pages = {024433},
  numpages = {12},
  year = {2025},
}

@Article{Cheong2024,
author={Cheong, Sang-Wook
and Huang, Fei-Ting},
title={Altermagnetism with non-collinear spins},
journal={npj Quantum Mater.},
year={2024},
month={Jan},
day={22},
volume={9},
number={1},
pages={13},
}

@article{sheoran2025spontaneousanomaloushalleffect,
  title={Spontaneous anomalous {H}all effect in two-dimensional altermagnets},
  author={Sheoran, Sajjan and Dev, Pratibha},
  journal={Phys. Rev. B},
  volume={111},
  number={18},
  pages={184407},
  year={2025},
  publisher={APS}
}

@article{PhysRevMaterials.8.L051401,
  title = {Nonrelativistic spin splittings and altermagnetism in twisted bilayers of centrosymmetric antiferromagnets},
  author = {Sheoran, Sajjan and Bhattacharya, Saswata},
  journal = {Phys. Rev. Mater.},
  volume = {8},
  issue = {5},
  pages = {L051401},
  numpages = {8},
  year = {2024},
  month = {May},
  publisher = {American Physical Society},
}

@article{Takahashi2025,
  title={Symmetry and minimal {H}amiltonian of nonsymmorphic collinear antiferromagnet {MnTe}},
  author={Takahashi, Koichiro and Huang, Hong-Fei and Yu, Jie-Xiang and Zang, Jiadong},
  journal={npj Quantum Mater.},
  volume={10},
  number={1},
  pages={1--11},
  year={2025},
  publisher={Nature Publishing Group}
}

@Article{Yuan2023,
author={Yuan, Lin-Ding
and Zhang, Xiuwen
and Acosta, Carlos Mera
and Zunger, Alex},
title={Uncovering spin-orbit coupling-independent hidden spin polarization of energy bands in antiferromagnets},
journal={Nat. Commun.},
year={2023},
month={Aug},
day={31},
volume={14},
number={1},
pages={5301},
}

@Article{Xiao2021,
author={Xiao, Rui-Chun
and Shao, Ding-Fu
and Li, Yu-Hang
and Jiang, Hua},
title={Spin photogalvanic effect in two-dimensional collinear antiferromagnets},
journal={npj Quantum Mater.},
year={2021},
month={Apr},
day={06},
volume={6},
number={1},
pages={35},
}

@article{doi:10.1126/sciadv.adn3662,
author = {Grgur Palle  and Risto Ojajärvi  and Rafael M. Fernandes  and Jörg Schmalian },
title = {Superconductivity due to fluctuating loop currents},
journal = {Sci. Adv.},
volume = {10},
number = {24},
pages = {eadn3662},
year = {2024},
}

@article{sobral2024fractionalizedaltermagnetsneighboringaltermagnetic,
	title = {Fractionalized altermagnets: From neighboring and altermagnetic spin liquids to spin-symmetric band splitting},
	author = {Sobral, Jo{\~a}o Augusto and Mandal, Subrata and Scheurer, Mathias S.},
	journal = {Phys. Rev. Res.},
	volume = {7},
	issue = {2},
	pages = {023152},
	numpages = {14},
	year = {2025},
	month = {May},
	publisher = {American Physical Society},
}

@article{autieri2312staggered,
  title = {Staggered {D}zyaloshinskii-{M}oriya interaction inducing weak ferromagnetism in centrosymmetric altermagnets and weak ferrimagnetism in noncentrosymmetric altermagnets},
  author = {Autieri, Carmine and Sattigeri, Raghottam M. and Cuono, Giuseppe and Fakhredine, Amar},
  journal = {Phys. Rev. B},
  volume = {111},
  issue = {5},
  pages = {054442},
  numpages = {17},
  year = {2025},
  month = {Feb},
  publisher = {American Physical Society},
}

@article{vasp_1993,
	title = {Ab initio molecular dynamics for liquid metals},
	author = {Kresse, G. and Hafner, J.},
	journal = {Phys. Rev. B},
	volume = {47},
	issue = {1},
	pages = {558--561},
	numpages = {0},
	year = {1993},
	month = {Jan},
	publisher = {American Physical Society},
}

@article{vasp_1996,
	title = {Efficient iterative schemes for ab initio total-energy calculations using a plane-wave basis set},
	author = {Kresse, G. and Furthm\"uller, J.},
	journal = {Phys. Rev. B},
	volume = {54},
	issue = {16},
	pages = {11169--11186},
	numpages = {0},
	year = {1996},
	month = {Oct},
	publisher = {American Physical Society},
}

@article{allen_1977,
	title = {Optical properties and electronic structure of crossroads material {MnTe}},
	journal = {Solid State Commun.},
	volume = {24},
	number = {5},
	pages = {367-370},
	year = {1977},
	author = {J.W. Allen and G. Lucovsky and J.C. Mikkelsen},
}

@article{cheong2025altermagnetismclassification,
author={Cheong, Sang-Wook
and Huang, Fei-Ting},
title={Altermagnetism classification},
journal={npj Quantum Mater.},
year={2025},
month={Apr},
day={12},
volume={10},
number={1},
pages={38},
}

@article{roig2024quasisymmetryconstrainedspinferromagnetism,
	title = {Quasisymmetry-Constrained Spin Ferromagnetism in Altermagnets},
	author = {Roig, Merc\`e and Yu, Yue and Ekman, Rune C. and Kreisel, Andreas and Andersen, Brian M. and Agterberg, Daniel F.},
	journal = {Phys. Rev. Lett.},
	volume = {135},
	issue = {1},
	pages = {016703},
	numpages = {8},
	year = {2025},
	month = {Jul},
	publisher = {American Physical Society},
}

@article{krempasky2024altermagnetic,
	title={Altermagnetic lifting of {K}ramers spin degeneracy},
	author={Krempask{\'y}, Juraj and {\v{S}}mejkal, L and D’souza, SW and Hajlaoui, M and Springholz, G and Uhl{\'\i}{\v{r}}ov{\'a}, K and Alarab, F and Constantinou, PC and Strocov, V and Usanov, D and others},
	journal={Nature},
	volume={626},
	number={7999},
	pages={517--522},
	year={2024},
	publisher={Nature Publishing Group UK London}
}

@article{xiao_2010,
	title = {{B}erry phase effects on electronic properties},
	author = {Xiao, Di and Chang, Ming-Che and Niu, Qian},
	journal = {Rev. Mod. Phys.},
	volume = {82},
	issue = {3},
	pages = {1959--2007},
	numpages = {0},
	year = {2010},
	month = {Jul},
	publisher = {American Physical Society},
}

@article{mcclarty_2024,
	title = {{L}andau Theory of Altermagnetism},
	author = {McClarty, Paul A. and Rau, Jeffrey G.},
	journal = {Phys. Rev. Lett.},
	volume = {132},
	issue = {17},
	pages = {176702},
	numpages = {8},
	year = {2024},
	month = {Apr},
	publisher = {American Physical Society},
}

@article{moriya_1960,
	title = {Anisotropic Superexchange Interaction and Weak Ferromagnetism},
	author = {Moriya, T\^oru},
	journal = {Phys. Rev.},
	volume = {120},
	issue = {1},
	pages = {91--98},
	numpages = {0},
	year = {1960},
	month = {Oct},
	publisher = {American Physical Society},
}

@article{johansson2024theory,
	title={Theory of spin and orbital {E}delstein effects},
	author={Johansson, Annika},
	journal={J. Condens. Matter Phys.},
	volume={36},
	number={42},
	pages={423002},
	year={2024},
	publisher={IOP Publishing}
}

@article{burgos_2024,
author = {Rhonald Burgos Atencia, Amit Agarwal and Dimitrie Culcer},
title = {Orbital angular momentum of {B}loch electrons: equilibrium formulation, magneto-electric phenomena, and the orbital {H}all effect},
journal = {Adv. Phys.: X},
volume = {9},
number = {1},
pages = {2371972},
year = {2024},
publisher = {Taylor \& Francis},
}

@article{xiao_2005,
	title = {{B}erry Phase Correction to Electron Density of States in Solids},
	author = {Xiao, Di and Shi, Junren and Niu, Qian},
	journal = {Phys. Rev. Lett.},
	volume = {95},
	issue = {13},
	pages = {137204},
	numpages = {4},
	year = {2005},
	month = {Sep},
	publisher = {American Physical Society},
}

@article{thonhauser_2005,
	title = {Orbital Magnetization in Periodic Insulators},
	author = {Thonhauser, T. and Ceresoli, Davide and Vanderbilt, David and Resta, R.},
	journal = {Phys. Rev. Lett.},
	volume = {95},
	issue = {13},
	pages = {137205},
	numpages = {4},
	year = {2005},
	month = {Sep},
	publisher = {American Physical Society},
}

@article{ceresoli_2006,
	title = {Orbital magnetization in crystalline solids: Multi-band insulators, {C}hern insulators, and metals},
	author = {Ceresoli, Davide and Thonhauser, T. and Vanderbilt, David and Resta, R.},
	journal = {Phys. Rev. B},
	volume = {74},
	issue = {2},
	pages = {024408},
	numpages = {13},
	year = {2006},
	month = {Jul},
	publisher = {American Physical Society},
}

@article{vanderbilt_2012,
	title = {{W}annier-based calculation of the orbital magnetization in crystals},
	author = {Lopez, M. G. and Vanderbilt, David and Thonhauser, T. and Souza, Ivo},
	journal = {Phys. Rev. B},
	volume = {85},
	issue = {1},
	pages = {014435},
	numpages = {12},
	year = {2012},
	month = {Jan},
	publisher = {American Physical Society},
}

@article{Bugajewski:2025_prb,
  title = {Theory of bound magnetic polarons in cubic and uniaxial antiferromagnets},
  author = {Bugajewski, Dawid and Autieri, Carmine and Dietl, Tomasz},
  journal = {Phys. Rev. B},
  volume = {112},
  issue = {14},
  pages = {L140403},
  numpages = {6},
  year = {2025},
}

@article{hanke_2016,
	title = {Role of {B}erry phase theory for describing orbital magnetism: From magnetic heterostructures to topological orbital ferromagnets},
	author = {Hanke, J.-P. and Freimuth, F. and Nandy, A. K. and Zhang, H. and Bl\"ugel, S. and Mokrousov, Y.},
	journal = {Phys. Rev. B},
	volume = {94},
	issue = {12},
	pages = {121114},
	numpages = {5},
	year = {2016},
	month = {Sep},
	publisher = {American Physical Society},
}

@misc{solovyev2025altermagnetismweakferromagnetism,
      title={Altermagnetism and Weak Ferromagnetism}, 
      author={I. V. Solovyev and S. A. Nikolaev and A. Tanaka},
      year={2025},
      eprint={2503.23735},
      archivePrefix={arXiv},
}

@article{meyer_1961,
	author = {Meyer, A. J. P. and Asch, G.},
	title = {Experimental g$^{\prime}$ and g Values of {F}e, {C}o, {N}i, and Their Alloys},
	journal = {J. Appl. Phys.},
	volume = {32},
	number = {3},
	pages = {S330-S333},
	year = {1961},
	month = {03},
}

@article{kresse_1996_2,
	title = {Efficiency of ab-initio total energy calculations for metals and semiconductors using a plane-wave basis set},
	journal = {Comp. Mat. Sci.},
	volume = {6},
	number = {1},
	pages = {15-50},
	year = {1996},
	author = {G. Kresse and J. Furthmüller},
}

@article{liechtenstein_1995,
	title = {Density-functional theory and strong interactions: Orbital ordering in {M}ott-{H}ubbard insulators},
	author = {Liechtenstein, A. I. and Anisimov, V. I. and Zaanen, J.},
	journal = {Phys. Rev. B},
	volume = {52},
	issue = {8},
	pages = {R5467--R5470},
	numpages = {0},
	year = {1995},
	month = {Aug},
	publisher = {American Physical Society},
}

@article{perdew_1996,
	title = "{Generalized gradient approximation made simple}",
	author = {Perdew, J. P. and Burke, K. and Ernzerhof, M.},
	journal = {Phys. Rev. Lett.},
	volume = {77},
	issue = {18},
	pages = {3865--3868},
	numpages = {0},
	year = {1996},
	month = {Oct},
	publisher = {American Physical Society},
}

@Article{Amin2024,
author={Amin, O. J.
and Dal Din, A.
and Golias, E.
and Niu, Y.
and Zakharov, A.
and Fromage, S. C.
and Fields, C. J. B.
and Heywood, S. L.
and Cousins, R. B.
and Maccherozzi, F.
and Krempask{\'y}, J.   
and others},
title={Nanoscale imaging and control of altermagnetism in {MnTe}},
journal={Nature},
year={2024},
month={Dec},
day={01},
volume={636},
number={8042},
pages={348-353},
}

@article{PhysRevLett.132.176701,
  title = {X-Ray Magnetic Circular Dichroism in Altermagnetic $\ensuremath{\alpha}$-{MnTe}},
  author = {Hariki, A. and Dal Din, A. and Amin, O. J. and Yamaguchi, T. and Badura, A. and Kriegner, D. and Edmonds, K. W. and Campion, R. P. and Wadley, P. and Backes, D. and others},
  journal = {Phys. Rev. Lett.},
  volume = {132},
  issue = {17},
  pages = {176701},
  numpages = {7},
  year = {2024},
  month = {Apr},
  publisher = {American Physical Society},
}

@article{PhysRevB.111.064401,
  title = {Spin quenching and transport by hidden {D}zyaloshinskii-{M}oriya interactions},
  author = {Ye, Xiyin and Cui, Qirui and Lin, Weiwei and Yu, Tao},
  journal = {Phys. Rev. B},
  volume = {111},
  issue = {6},
  pages = {064401},
  numpages = {11},
  year = {2025},
  month = {Feb},
  publisher = {American Physical Society},
}

@article{Jo2024gtensor,
  title = {Weak Ferromagnetism in Altermagnets from Alternating $g$-Tensor Anisotropy},
  author = {Jo, Daegeun and Go, Dongwook and Mokrousov, Yuriy and Oppeneer, Peter M. and Cheong, Sang-Wook and Lee, Hyun-Woo},
  journal = {Phys. Rev. Lett.},
  volume = {134},
  issue = {19},
  pages = {196703},
  numpages = {9},
  year = {2025},
  month = {May},
  publisher = {American Physical Society},
}

@article{PhysRevB.105.235142,
  title = {Orbital angular momentum driven anomalous {H}all effect},
  author = {Dowinton, Oliver and Bahramy, Mohammad Saeed},
  journal = {Phys. Rev. B},
  volume = {105},
  issue = {23},
  pages = {235142},
  numpages = {10},
  year = {2022},
  month = {Jun},
  publisher = {American Physical Society},
}

@article{tamang2024newlydiscoveredmagneticphase,
  title={Altermagnetism and altermagnets: A brief review},
  author={Tamang, Rupam and Gurung, Shivraj and Rai, Dibya Prakash and Brahimi, Samy and Lounis, Samir},
  journal={Magnetism},
  volume={5},
  number={3},
  pages={17},
  year={2025},
  publisher={MDPI}
}

@Article{Takagi2025,
author={Takagi, Rina
and Hirakida, Ryosuke
and Settai, Yuki
and Oiwa, Rikuto
and Takagi, Hirotaka
and Kitaori, Aki
and Yamauchi, Kensei
and Inoue, Hiroki
and Yamaura, Jun-ichi
and Nishio-Hamane, Daisuke
and others},
title={Spontaneous {H}all effect induced by collinear antiferromagnetic order at room temperature},
journal={Nat. Mater.},
year={2025},
month={Jan},
day={01},
volume={24},
number={1},
pages={63-68},
}

@article{PhysRevB.103.115209,
  title = {Momentum-resolved spin splitting in {M}n-doped trivial {CdTe} and topological {HgTe} semiconductors},
  author = {Autieri, Carmine and \ifmmode \acute{S}\else \'{S}\fi{}liwa, Cezary and Islam, Rajibul and Cuono, Giuseppe and Dietl, Tomasz},
  journal = {Phys. Rev. B},
  volume = {103},
  issue = {11},
  pages = {115209},
  numpages = {18},
  year = {2021},
  month = {Mar},
  publisher = {American Physical Society},
}

@article{takegami2025circulardichroismresonantinelastic,
  title = {Circular dichroism in resonant inelastic {X}-ray Scattering: Probing altermagnetic domains in {MnTe}},
  author = {D. Takegami and T. Aoyama and T. Okauchi and T. Yamaguchi and S. Tippireddy and S. Agrestini and M. García-Fernández and T. Mizokawa and K. Ohgushi and Ke-Jin Zhou and J. Chaloupka and J. Kuneš and A. Hariki and H. Suzuki},
  journal = {Phys. Rev. Lett.},
  volume={135},
  number={19},
  pages={196502},
  year={2025},
  publisher={APS}
}

@article{PhysRevMaterials.8.104407,
  title = {Interplay of altermagnetism and pressure in hexagonal and orthorhombic {MnTe}},
  author = {Devaraj, Nayana and Bose, Anumita and Narayan, Awadhesh},
  journal = {Phys. Rev. Mater.},
  volume = {8},
  issue = {10},
  pages = {104407},
  numpages = {15},
  year = {2024},
  month = {Oct},
  publisher = {American Physical Society},
}

@article{pizzi2020,
	year = 2020,
	month = {jan},
	publisher = {{IOP} Publishing},
	volume = {32},
	number = {16},
	pages = {165902},
	author = {Giovanni Pizzi and Valerio Vitale and Ryotaro Arita and Stefan Blügel and Frank Freimuth and Guillaume G{\'{e}}ranton and Marco Gibertini and Dominik Gresch and Charles Johnson and Takashi Koretsune and others},
	title = {Wannier90 as a community code: new features and applications},
	journal = {J. Condens. Matter Phys.}
}

@article{antropov_2014,
	title = {Magnetic anisotropic effects and electronic correlations in {MnBi} ferromagnet},
	author = {Antropov, V. P. and Antonov, V. N. and Bekenov, L. V. and Kutepov, A. and Kotliar, G.},
	journal = {Phys. Rev. B},
	volume = {90},
	issue = {5},
	pages = {054404},
	numpages = {18},
	year = {2014},
	month = {Aug},
	publisher = {American Physical Society},
}

@article{paoflow_2018,
	title = "{PAOFLOW: A utility to construct and operate on ab initio Hamiltonians from the projections of electronic wavefunctions on atomic orbital bases, including characterization of topological materials}",
	journal = {Comp. Mat. Sci.},
	volume = {143},
	pages = {462-472},
	year = {2018},
	author = {Marco {Buongiorno Nardelli} and Frank T. Cerasoli and Marcio Costa and Stefano Curtarolo and Riccardo {De Gennaro} and Marco Fornari and Laalitha Liyanage and Andrew R. Supka and Haihang Wang}
}

@article{paoflow_2021,
	title = "{Advanced modeling of materials with PAOFLOW 2.0: new features and software design}",
	journal = {Comp. Mat. Sci.},
	volume = {200},
	pages = {110828},
	year = {2021},
	author = {Frank T. Cerasoli and Andrew R. Supka and Anooja Jayaraj and Marcio Costa and Ilaria Siloi and Jagoda S{\l}awi{\'n}ska and Stefano Curtarolo and Marco Fornari and Davide Ceresoli and Marco {Buongiorno Nardelli}}
}

@article{kubler_1981,
	title = {Magnetic moments of ferromagnetic and antiferromagnetic bcc and fcc iron},
	journal = {Phys. Lett. A},
	volume = {81},
	number = {1},
	pages = {81-83},
	year = {1981},
	author = {J. Kübler},
}

@article{smejkal_2022,
	title = {Beyond Conventional Ferromagnetism and Antiferromagnetism: A Phase with Nonrelativistic Spin and Crystal Rotation Symmetry},
	author = {\ifmmode \check{S}\else \v{S}\fi{}mejkal, Libor and Sinova, Jairo and Jungwirth, Tomas},
	journal = {Phys. Rev. X},
	volume = {12},
	issue = {3},
	pages = {031042},
	numpages = {16},
	year = {2022},
	month = {Sep},
	publisher = {American Physical Society},
}

@article{smejkal_2022_2,
	title = {Emerging Research Landscape of Altermagnetism},
	author = {\ifmmode \check{S}\else \v{S}\fi{}mejkal, Libor and Sinova, Jairo and Jungwirth, Tomas},
	journal = {Phys. Rev. X},
	volume = {12},
	issue = {4},
	pages = {040501},
	numpages = {27},
	year = {2022},
	month = {Dec},
	publisher = {American Physical Society},
}

@article{smejkal_2022_3,
	title = {Giant and Tunneling Magnetoresistance in Unconventional Collinear Antiferromagnets with Nonrelativistic Spin-Momentum Coupling},
	author = {\ifmmode \check{S}\else \v{S}\fi{}mejkal, Libor and Hellenes, Anna Birk and Gonz\'alez-Hern\'andez, Rafael and Sinova, Jairo and Jungwirth, Tomas},
	journal = {Phys. Rev. X},
	volume = {12},
	issue = {1},
	pages = {011028},
	numpages = {11},
	year = {2022},
	month = {Feb},
	publisher = {American Physical Society},
}

@article{kluczyk_2024,
	title = {Coexistence of anomalous {H}all effect and weak magnetization in a nominally collinear antiferromagnet {MnTe}},
	  author = {Kluczyk, K. P. and Gas, K. and Grzybowski, M. J. and Skupi\'{n}ski, P. and Borysiewicz, M. A. and F\c{a}s, T. and Suffczy\'{n}ski, J. and Domagala, J. Z. and Grasza, K. and Mycielski, A. and others},
	journal = {Phys. Rev. B},
	volume = {110},
	issue = {15},
	pages = {155201},
	numpages = {12},
	year = {2024},
	month = {Oct},
	publisher = {American Physical Society},
}

@article{guo_2023,
	title = {Spin-split collinear antiferromagnets: A large-scale ab-initio study},
	journal = {Mater. Today Phys.},
	volume = {32},
	pages = {100991},
	year = {2023},
	author = {Yaqian Guo and Hui Liu and Oleg Janson and Ion Cosma Fulga and Jeroen {van den Brink} and Jorge I. Facio},
}

@article{cuono_2023,
	title = {Ab initio overestimation of the topological region in {Eu}-based compounds},
	author = {Cuono, Giuseppe and Sattigeri, Raghottam M. and Autieri, Carmine and Dietl, Tomasz},
	journal = {Phys. Rev. B},
	volume = {108},
	issue = {7},
	pages = {075150},
	numpages = {12},
	year = {2023},
	month = {Aug},
	publisher = {American Physical Society},
}

@article{gonzalez_2021,
	title = {Efficient Electrical Spin Splitter Based on Nonrelativistic Collinear Antiferromagnetism},
	author = {Gonz\'alez-Hern\'andez, Rafael and \ifmmode \check{S}\else \v{S}\fi{}mejkal, Libor and V\'yborn\'y, Karel and Yahagi, Yuta and Sinova, Jairo and Jungwirth, Tom\'a\ifmmode \check{s}\else \v{s}\fi{} and \ifmmode \check{Z}\else \v{Z}\fi{}elezn\'y, Jakub},
	journal = {Phys. Rev. Lett.},
	volume = {126},
	issue = {12},
	pages = {127701},
	numpages = {6},
	year = {2021},
	month = {Mar},
	publisher = {American Physical Society},
}

@article{ding_2021,
	title={Spin-neutral currents for spintronics},
	author={Shao, Ding-Fu and Zhang, Shu-Hui and Li, Ming and Eom, Chang-Beom and Tsymbal, Evgeny Y},
	journal={Nat. Commun.},
	volume={12},
	number={1},
	pages={7061},
	year={2021},
	publisher={Nature Publishing Group UK London}
}

@article{ding_2023,
	title = {N\'eel Spin Currents in Antiferromagnets},
	author = {Shao, Ding-Fu and Jiang, Yuan-Yuan and Ding, Jun and Zhang, Shu-Hui and Wang, Zi-An and Xiao, Rui-Chun and Gurung, Gautam and Lu, W. J. and Sun, Y. P. and Tsymbal, Evgeny Y.},
	journal = {Phys. Rev. Lett.},
	volume = {130},
	issue = {21},
	pages = {216702},
	numpages = {8},
	year = {2023},
	month = {May},
	publisher = {American Physical Society},
}

@article{zhou_2024,
	title = {Crystal Thermal Transport in Altermagnetic {RuO}$_{2}$},
	author = {Zhou, Xiaodong and Feng, Wanxiang and Zhang, Run-Wu and \ifmmode \check{S}\else \v{S}\fi{}mejkal, Libor and Sinova, Jairo and Mokrousov, Yuriy and Yao, Yugui},
	journal = {Phys. Rev. Lett.},
	volume = {132},
	issue = {5},
	pages = {056701},
	numpages = {7},
	year = {2024},
	month = {Jan},
	publisher = {American Physical Society},
}

@article{ouassou_2023,
	title = {dc {J}osephson Effect in Altermagnets},
	author = {Ouassou, Jabir Ali and Brataas, Arne and Linder, Jacob},
	journal = {Phys. Rev. Lett.},
	volume = {131},
	issue = {7},
	pages = {076003},
	numpages = {6},
	year = {2023},
	month = {Aug},
	publisher = {American Physical Society},
}

@article{dzyaloshinsky_1958,
	title = {A thermodynamic theory of ``weak” ferromagnetism of antiferromagnetics},
	journal = {J. Phys. Chem. Solids},
	volume = {4},
	number = {4},
	pages = {241-255},
	year = {1958},
	author = {I. Dzyaloshinsky},
}

@article{yin_2019,
  title = {Planar {H}all Effect in Antiferromagnetic {MnTe} Thin Films},
  author = {Yin, Gen and Yu, Jie-Xiang and Liu, Yizhou and Lake, Roger K. and Zang, Jiadong and Wang, Kang L.},
  journal = {Phys. Rev. Lett.},
  volume = {122},
  issue = {10},
  pages = {106602},
  numpages = {6},
  year = {2019},
  month = {Mar},
  publisher = {American Physical Society},
}

@article{belashchenko_2025,
	title = {Giant Strain-Induced Spin Splitting Effect in {MnTe}, a $g$-Wave Altermagnetic Semiconductor},
	author = {Belashchenko, K. D.},
	journal = {Phys. Rev. Lett.},
	volume = {134},
	issue = {8},
	pages = {086701},
	numpages = {6},
	year = {2025},
	month = {Feb},
	publisher = {American Physical Society},
}

@article{lee_2024,
	title = {Broken {K}ramers Degeneracy in Altermagnetic {MnTe}},
	author = {Lee, Suyoung and Lee, Sangjae and Jung, Saegyeol and Jung, Jiwon and Kim, Donghan and Lee, Yeonjae and Seok, Byeongjun and Kim, Jaeyoung and Park, Byeong Gyu and \ifmmode \check{S}\else \v{S}\fi{}mejkal, Libor and Kang, Chang-Jong and Kim, Changyoung},
	journal = {Phys. Rev. Lett.},
	volume = {132},
	issue = {3},
	pages = {036702},
	numpages = {7},
	year = {2024},
	month = {Jan},
	publisher = {American Physical Society},
}

@article{kriegner2016multiple,
	title={Multiple-stable anisotropic magnetoresistance memory in antiferromagnetic {MnTe}},
	author={Kriegner, Dominik and V{\`y}born{\`y}, K and Olejn{\'\i}k, K and Reichlov{\'a}, H and Nov{\'a}k, V and Marti, X and Gazquez, J and Saidl, V and N{\v{e}}mec, P and Volobuev, VV and others},
	journal={Nat. Commun.},
	volume={7},
	number={1},
	pages={11623},
	year={2016},
	publisher={Nature Publishing Group UK London}
}

@article{liu_2024,
	title = {Chiral Split Magnon in Altermagnetic {MnTe}},
	author = {Liu, Zheyuan and Ozeki, Makoto and Asai, Shinichiro and Itoh, Shinichi and Masuda, Takatsugu},
	journal = {Phys. Rev. Lett.},
	volume = {133},
	issue = {15},
	pages = {156702},
	numpages = {6},
	year = {2024},
	month = {Oct},
	publisher = {American Physical Society},
}

@article{mori2020reversible,
	title={Reversible displacive transformation in {MnTe} polymorphic semiconductor},
	author={Mori, Shunsuke and Hatayama, Shogo and Shuang, Yi and Ando, Daisuke and Sutou, Yuji},
	journal={Nat. Commun.},
	volume={11},
	number={1},
	pages={85},
	year={2020},
	publisher={Nature Publishing Group UK London}
}

@Article{GonzalezBetancourt2024,
	author={Gonzalez Betancourt, Ruben Dario
	and Zub{\'a}{\v{c}}, Jan
	and Geishendorf, Kevin
	and Ritzinger, Philipp
	and R{\r{u}}{\v{z}}i{\v{c}}kov{\'a}, Barbora
	and Kotte, Tommy
	and {\v{Z}}elezn{\'y}, Jakub
	and Olejn{\'i}k, Kamil
	and Springholz, Gunther
	and Büchner, Bernd
	and others},
	title={Anisotropic magnetoresistance in altermagnetic {MnTe}},
	journal={npj Spintronics},
	year={2024},
	month={Aug},
	day={13},
	volume={2},
	number={1},
	pages={45},
}

@article{mazurenko2005weak,
	title = {Weak ferromagnetism in antiferromagnets: $\alpha$-{Fe}$_{2}${O}$_{3}$ and {La}$_{2}${Cu}{O}$_{4}$},
	author = {Mazurenko, V. V. and Anisimov, V. I.},
	journal = {Phys. Rev. B},
	volume = {71},
	issue = {18},
	pages = {184434},
	numpages = {8},
	year = {2005},
	month = {May},
	publisher = {American Physical Society},
}

@article{Togo31122024,
author = {Atsushi Togo and Kohei Shinohara and Isao Tanaka},
title = {Spglib: a software library for crystal symmetry search},
journal = {STAM-M},
volume = {4},
number = {1},
pages = {2384822},
year = {2024},
publisher = {Taylor \& Francis},
}

@article{spglib2,
  title={Algorithms for magnetic symmetry operation search and identification of magnetic space group from magnetic crystal structure},
  author={Shinohara, Kohei and Togo, Atsushi and Tanaka, Isao},
  journal={Acta Crystallogr. A: Found. Crystallogr.},
  volume={79},
  number={5},
  pages={390--398},
  year={2023},
  publisher={International Union of Crystallography}
}

@article{PhysRevB.90.045202,
  title = {Orbital magnetization in dilute ferromagnetic semiconductors},
  author = {\ifmmode \acute{S}\else \'{S}\fi{}liwa, Cezary and Dietl, Tomasz},
  journal = {Phys. Rev. B},
  volume = {90},
  issue = {4},
  pages = {045202},
  numpages = {5},
  year = {2014},
  month = {Jul},
  publisher = {American Physical Society}
}

@article{srdjan,
year = {2024},
volume = {11},
number = {3},
pages = {035025},
author = {Milivojević, Marko and Orozović, Marko and Picozzi, Silvia and Gmitra, Martin and Stavrić, Srđan},
title = {Interplay of altermagnetism and weak ferromagnetism in two-dimensional {RuF}$_{4}$},
journal = {2D Materials},
}

@article{liu2025different,
  title={Different facets of unconventional magnetism},
  author={Liu, Qihang and Dai, Xi and Bl{\"u}gel, Stefan},
  journal={Nat. Phys},
  volume={21},
  number={3},
  pages={329--331},
  year={2025},
}

@article{Cracknell1969,
  title={Group theory and magnetic phenomena in solids},
  author={Cracknell, Arthur Philip},
  journal={Rep. Prog. Phys.},
  volume={32},
  number={2},
  pages={633},
  year={1969},
  publisher={IOP Publishing}
}

@article{koretsune_2018,
author = {Koretsune, Takashi and Kikuchi, Toru and Arita, Ryotaro},
title = {First-Principles Evaluation of the {D}zyaloshinskii–{M}oriya Interaction},
journal = {J. Phys. Soc. Jpn.},
volume = {87},
number = {4},
pages = {041011},
year = {2018},
}

@article{beutier_2017,
  title = {Band Filling Control of the {D}zyaloshinskii-{M}oriya Interaction in Weakly Ferromagnetic Insulators},
  author = {Beutier, G. and Collins, S. P. and Dimitrova, O. V. and Dmitrienko, V. E. and Katsnelson, M. I. and Kvashnin, Y. O. and Lichtenstein, A. I. and Mazurenko, V. V. and Nisbet, A. G. A. and Ovchinnikova, E. N. and Pincini, D.},
  journal = {Phys. Rev. Lett.},
  volume = {119},
  issue = {16},
  pages = {167201},
  numpages = {6},
  year = {2017},
  month = {Oct},
  publisher = {American Physical Society},
}

@article{qe1,
  author={Paolo Giannozzi and Stefano Baroni and Nicola Bonini and Matteo Calandra and Roberto Car and Carlo Cavazzoni and Davide
Ceresoli and Guido L Chiarotti and Matteo Cococcioni and Ismaila Dabo and Andrea Dal Corso and Stefano de
Gironcoli and others},
  title={{QUANTUM ESPRESSO}: a modular and open-source software project for quantum simulations of materials},
  journal={J. Phys. Condens. Matter},
  volume={21},
  number={39},
  pages={395502},
  year={2009},
}

@article{qe2,
author = {Giannozzi, P and Andreussi, O and Brumme, T and Bunau, O and {Buongiorno Nardelli}, M and Calandra, M and Car, R and Cavazzoni, C and Ceresoli, D and Cococcioni, M and others},
title = {Advanced capabilities for materials modelling with {Q}uantum {ESPRESSO}},
journal = {J. Phys. Condens. Matter},
year = {2017},
volume = {29},
number = {46},
pages = {465901--31},
}

@ARTICLE{Li2025_Weyl,
author    = {Li, Cong and Hu, Mengli and Li, Zhilin and Wang, Yang and Chen, Wanyu and Thiagarajan, Balasubramanian and Leandersson, Mats and Polley, Craig and Kim, Timur and Liu, Hui and others},
  title     = {Topological Weyl altermagnetism in {CrSb}},
  journal   = {Communications Physics},
  publisher = {Springer Science and Business Media LLC},
  volume    = {8},
  number    = {1},
  month     = {July},
  year      = {2025},
  language  = {English}
}

\end{document}